\documentclass[preprintnumbers,notitlepage,a4paper,aps,prd,onecolumn,superscriptaddress,nofootinbib,groupedaddress]{revtex4}

\usepackage{amsmath}
\usepackage{amsfonts,color}
\usepackage{amsthm}
\usepackage{pdfpages}
\usepackage{empheq}
\usepackage{amssymb,float}
\usepackage{booktabs,multirow}
\usepackage{titlesec}
\usepackage{hyphenat}
\usepackage{ragged2e}
\usepackage{amsmath}
\usepackage{accents}
\usepackage[utf8]{inputenc}
\usepackage{color}
\usepackage{hyperref}
\usepackage{enumitem}
\usepackage{tikz}
\usepackage{comment}
\usepackage{ulem}
\usepackage{cancel}
\usetikzlibrary{shapes.geometric}
\usetikzlibrary{arrows.meta,arrows}

\titleformat{\paragraph}
{\normalfont\normalsize\bfseries}{\theparagraph}{1em}{}
\titlespacing*{\paragraph}
{0pt}{3.25ex plus 1ex minus .2ex}{1.5ex plus .2ex}

\allowdisplaybreaks
\hypersetup{
    colorlinks=true,
    linkcolor=blue,
    filecolor=magenta,      
    citecolor=blue
}

\allowdisplaybreaks[4] 
\begin{document}

\preprint{KEK-TH-2760, KEK-Cosmo-0393}

\title{
Cosmological Perturbation in New General Relativity: \\
Propagating mode from the violation of local Lorentz invariance
}

\author{Kyosuke Tomonari}
\email[]{ktomonari.phys@gmail.com}
\affiliation{Institute of Astrophysics, Central China Normal University, Wuhan 430079, China}
\affiliation{Interfaculty Initiative in Information Studies, Graduate School of Interdisciplinary Information Studies, The University of Tokyo. 7-3-1 Hongo, Bunkyo-ku, Tokyo 113-0033, Japan}

\author{Taishi~Katsuragawa}
\email[
]{taishi@ccnu.edu.cn}
\affiliation{Institute of Astrophysics, Central China Normal University, Wuhan 430079, China}

\author{Shin'ichi Nojiri}
\email[]{nojiri@nagoya-u.jp}
\affiliation{
KEK Theory Center, Institute of Particle and Nuclear Studies, 
High Energy Accelerator Research Organization (KEK), Oho 1-1, Tsukuba, Ibaraki 305-0801, Japan
}
\affiliation{
Kobayashi-Maskawa Institute for the Origin of Particles and the Universe, 
Nagoya University, Nagoya 464-8602, Japan
}

\begin{abstract}
We investigate the propagating modes of New General Relativity (NGR) in second-order linear perturbations in the Lagrangian density (first-order in field equations). 
The Dirac-Bergmann analysis has revealed a violation of local Lorentz invariance in NGR. 
We review the recent status of NGR, considering the results of its Dirac-Bergmann analysis. 
We then reconsider the vierbein perturbation framework and identify the origin of each perturbation field in the vierbein field components. 
This identification is mandatory for adequately fixing gauges while guaranteeing consistency with the invariance ensured by the Dirac-Bergmann analysis. 
We find that the spatially flat gauge is adequate for analyzing a theory with the violation of local Lorentz invariance.
Based on the established vierbein perturbative framework, introducing a real scalar field as matter, we perform a second-order perturbative analysis of NGR with respect to tensor, scalar, pseudo-scalar, and vector and pseudo-vector modes. 
We reveal the possible propagating modes of each type of NGR. 
In particular, we find that Type 3 has stable five propagating modes, \textit{i.e.}, tensor, scalar, and vector modes, compared to five non-linear degrees of freedom, which results in its Dirac-Bergmann analysis;
The linear perturbation theory of Type 3 is preferable for applications to cosmology. 
Finally, we discuss our results in comparison to previous related work and conclude this study.
\end{abstract}

\maketitle

\section{\label{01}Introduction}

New General Relativity (NGR) is an extension of Teleparallel Equivalent to General Relativity (TEGR) with three free parameters, in which torsion plays the primary role in describing gravity in a parity-preserving manner~\cite{Einstein1928, Hayashi:1979qx}. 
NGR provides richer degrees of freedom (DOF) compared to TEGR~\cite{Bahamonde:2021gfp}, and this abundance has the potential to elucidate issues in cosmology such as dark energy~\cite{SupernovaCosmologyProject:1998vns, SupernovaSearchTeam:1998fmf, Planck:2018nkj}, dark matter~\cite{Freese:2008cz, Billard:2021uyg, Planck:2018nkj}, and tensions in cosmological parameters~\cite{Planck:2018vyg, H0LiCOW:2019pvv, Schoneberg:2022ggi, Riess:2019cxk, ACT:2023kun}. 
One can check the present status of the tensions in Ref.~\cite{CosmoVerseNetwork:2025alb}.
To investigate these phenomenological issues, it is essential to clarify the nature of DOFs in NGR from the viewpoint of cosmological perturbations based on a constraint system. 
Recently, one of the authors has revealed the constraint structure and counted the DOFs of NGR~\cite{Tomonari:2024ybs, Tomonari:2024lpv}.
However, whereas propagating modes around Minkowski spacetime have been well investigated in previous works~\cite{VanNieuwenhuizen:1973fi, Kuhfuss:1986rb, Golovnev:2023ddv, Mikura:2023ruz, Bahamonde:2024zkb}, cosmological perturbations of NGR have not yet been sufficiently clarified~\cite{Golovnev:2023jnc}. 

Cosmological perturbations for NGR are special in the sense that NGR no longer satisfies the local Lorentz Invariance (LI)~\cite{Tomonari:2024lpv, Tomonari:2024ybs}. 
This violation of local LI, in turn, can produce new propagating modes that do not appear in conventional theories such as GR and $f(R)$ gravity. 
Despite this, the standard perturbation theory typically includes only metric components~\cite{Malik:2008im}, which correspond to the symmetric part of (co-)vierbein field components. 
To properly account for these additional propagating modes, it becomes necessary to incorporate the anti-symmetric part of the (co-)vierbein field into the perturbative framework, where the local Lorentz transformation operates. 

Historically, $f(T)$-gravity has encountered the same issues as NGR. 
The first result on Dirac-Bergmann (DB) analysis~\cite{BergmannBrunings1949, Bergmann:1949zz, Bergmann1950, Dirac:1950pj, Anderson:1951ta, Dirac:1958sc, Dirac:1958sq}, on the one hand, appeared in Ref.~\cite{Li:2011rn}. 
In this pioneering work, the authors implied the violation of local LI by stating that the first-class constraints corresponding to the local LI turn into second-class constraints. 
A closed algebra of first-class constraints in Poisson bracket forms a gauge symmetry~\cite{Sugano:1982bm, Sugano:1986xb, Sugano:1989rq, Sugano:1991ir, Sugano:1991ke, Sugano:1991kd}. 
Thus, the result means that the LI is lost, at least as a local invariance. 
In this point, see also Refs.~\cite{Blagojevic:2000pi, Blagojevic:2000qs, Li:2011rn, Ferraro:2018tpu, Blagojevic:2020dyq, Blixt:2020ekl, Blagojevic:2023fys} for details. 
On the other hand, a vierbein perturbation theory has been established~\cite{Dent:2010nbw, Chen:2010va, Wu:2012hs, Izumi:2012qj}. 
The authors in Refs.~\cite{Dent:2010nbw, Chen:2010va, Wu:2012hs} incorporated the anti-symmetric part of vierbein components into the perturbations, thereby enabling us to consider a perturbation theory with the violation of local LI. 
The authors in Ref.~\cite{Izumi:2012qj} completed the perturbative framework taking into account the inclusion of the pseudo-vector mode. 

Recently, cosmological perturbation on NGR has been performed using conformal transformations based on the results on Minkowski background spacetime~\cite{Golovnev:2023ddv, Golovnev:2023jnc}. 
A conformal Newtonian gauge is imposed to investigate the propagation of each perturbative mode. 
However, there are still concerns in this analysis for the following reasons.
1) The conventional perturbative framework is applied~\cite{Golovnev:2018wbh}, suggesting that the relationship between each perturbative variable and vierbein component remains unclear. 
In particular, in Ref.~\cite{Golovnev:2018wbh}, the perturbations are introduced without explicit derivations, suggesting that we should identify each origin of the perturbation field. 
2) The perturbative framework in this work does not take into account pioneering works~\cite{Dent:2010nbw, Chen:2010va, Wu:2012hs, Izumi:2012qj}, particularly in Ref.~\cite{Izumi:2012qj}. 
The consistency with Refs.~\cite{Golovnev:2018wbh, Golovnev:2023ddv, Golovnev:2023jnc} should be investigated.
3) Refs.~\cite{Golovnev:2018wbh, Golovnev:2023ddv, Golovnev:2023jnc} do not reflect the result of the DB analysis in NGR~\cite{Tomonari:2024lpv, Tomonari:2024ybs}, meaning that there may remain a doubt that the gauge fixing is not appropriate;
If this is the case, the propagating modes accounted for in this work could be insufficient.
4) The metric and torsion perturbations are calculated at first-order levels, implying that consideration of higher-order contributions in these variables may raise additional propagating modes.

To address these issues, we analyze the cosmological perturbation and reveal possible propagating modes in NGR.
We perform the perturbative expansion of the Lagrangian density of NGR up to the second order around the flat Friedmann-Lema\^{i}tre-Robertson-Walker (FLRW) spacetime.
We introduce a scalar field as a matter field to realize the flat FLRW background; otherwise, the scale factor is constant, and our analysis is always reduced to existing analysis~\cite{Bahamonde:2024zkb, Golovnev:2023ddv} in the Minkowski background up to the scaling of the spatial coordinates. 
We investigate the propagation of perturbation modes and discuss the number of propagating modes according to the known classification of the three parameters in NGR.
We summarize our results on the number of propagating modes in the Table.~\ref{Table:CosmoPertOfNGR}

The organization of this paper is as follows. 
In Sec.~\ref{02}, we review NGR from the point of view of DB analysis. 
In Sec.~\ref{03}, we reconstruct cosmological perturbations to clarify the origin of each perturbation in vierbein field components. 
In Sec.~\ref{04}, we derive the background equations of NGR incorporating a real scalar field as matter. 
In Sec.~\ref{05}, we investigate the propagating modes of tensor, (pseudo-)scalar, and (pseudo-)vector modes in each type of NGR. 
Finally, in Sec.~\ref{06}, we summarize and conclude this work. 

Throughout this paper, we use the unit of $c^{4} / 16 \pi G = 1\,$. 
We denote by greek letters, $\alpha\,,\beta\,,\gamma\,,\cdots\,,\mu\,,\nu\,,\rho\,,\cdots\,$,  spacetime indices, and by capital latin letters, $A\,,B\,,C\,,\cdots\,,I\,,J\,,K\,,\cdots$, internal-space indices, and by small latin letters, $a\,,b\,,c\,,\cdots$ and $i\,,j\,,k\,,\cdots\,$, internal-space spatial indices and spacetime spatial indices, respectively. 
We express the covariant derivative with respect to the Levi-Civita connection as $\overset{\circ}{\nabla}$ and distinguish it from the covariant derivative with respect to an affine connection as $\nabla$. 
For all calculations in this work, we utilized Cadabra~\cite{Peeters:2007wn}, a free and excellent calculator.

\section{\label{02}New General Relativity}

\subsection{\label{02:01}Fundamental ingredients}

NGR consists of a (co-)vierbein field and a connection 1-form in the internal-space formulation. 
In this work, we consider NGR based on the Weitzenb\"{o}ck gauge in four-dimensional spacetime. 
This gauge restricts an affine connection to the Weitzenb\"{o}ck connection given as follows:
\footnote{
We note that in TEGR the absence of spin connection gives rise to a boundary term when we perform a local Lorentz transformation to the action integral. 
This means that TEGR satisfies the local LI. 
}
\begin{equation}
\label{Weitzenboech connection}
    \overset{\mathrm{w}}{\Gamma}{}^{\rho}_{\mu\nu} 
    = 
    e_{I}{}^{\rho}\,\partial_{\mu}\,e^{I}{}_{\nu}
    \,,
\end{equation}
where $e_{I}{}^{\mu}$ and $e^{I}{}_{\mu}$ are the vierbein and the co-vierbein field (components), respectively. 
Here, these quantities are related by $e_{I}{}^{\mu} e^{I}{}_{\nu} = \delta^{\mu}{}_{\nu}\,$, or equivalently, $e_{I}{}^{\mu} e^{J}{}_{\mu} = \delta^{J}{}_{I}\,$.
Thus, NGR needs only the (co-)vierbein to formulate it.
Using this connection, the torsion tensor turns out to be
\begin{equation}
    \overset{\mathrm{w}}{T}{}^{\rho}{}_{\mu\nu} 
    = 
    2\,\overset{\mathrm{w}}{\Gamma}{}^{\rho}_{[\mu\nu]} 
    = 
    e_{I}{}^{\rho}\,\left(\,\partial_{\mu}e^{I}{}_{\nu} 
    - \partial_{\nu}e^{I}{}_{\mu}\,\right)
    \,.
\label{torsion in W-gauge}
\end{equation}
Apparently, Eq.~\eqref{Weitzenboech connection} does not satisfy the local LI. 
The same statement holds for Eq.~\eqref{torsion in W-gauge}. 
These properties are an implication of the violation in NGR.\footnote{
In TEGR (Type 6 of NGR), a local Lorentz transformation of the Lagrangian provides a boundary term in the action integral. 
Thus, the theory guarantees the invariance only when the action contains boundary terms. 
In contrast, this is generally not the case in NGR except for Type 6 (TEGR).
}
In detail, see Sec.~II-A in Ref.~\cite{Tomonari:2024ybs} and the original work Ref.~\cite{Blixt:2018znp}. 

To verify this violation in NGR, there are two methods: 
1) Derive the field equation with respect to the tetrad field components and check the anti-symmetric part of the equation. 
If the anti-symmetric part of the equation identically holds, the theory satisfies the local LI. 
If this is not the case, extra DOFs would arise up to six; 
2) Perform the DB analysis~\cite{Bergmann:1949zz, BergmannBrunings1949, Dirac:1950pj, Bergmann1950, Anderson:1951ta, Dirac:1958sq, Dirac:1958sc} and check the Poisson Bracket algebra (PB algebra) of constraints in the vierbein sector. 
If the algebra shows the Lie algebra of $SO(1\,,3)$, the given theory satisfies the local LI. 
If this is not the case, the theory violates the invariance and gives rise to extra DOFs up to six.\footnote{
In NGR, since the diffeomorphism invariance holds, the upper bound of non-linear DOFs is $(16\times2 - 8\times2)/2 = 8\,$. 
The two DOFs of eight are none other than the DOFs describing gravity (tensor modes).
The remaining six DOFs are ascribed to the violation of the local Lorentz invariance.
See also Refs.~\cite{Bahamonde:2024zkb,Tomonari:2024ybs,Tomonari:2024lpv} in this point.
}

Constraint systems are classified into two types: regular and irregular systems~\cite{Dirac:1950pj, Dirac:1958sc, Dirac:1958sq, Miskovic:2003ex}.
For regular systems, based on the DB analysis, we can calculate the non-linear DOFs by using the following formula: \textit{nonlinear DOFs $ = ($phase space dimension $-$ $2\times\#$ of first-class constraint(s) $-$ $\#$ of second-class constraints $)/2$}, where the symbol $\#$ denotes ``number''.
Here, the regularity of a system is defined by that the first-order variation of every constraint is expressed by a linear combination of constraints existing in the theory.
In the constraint Hamiltonian formulation, a closed PB algebra of first-class constraints provides a gauge symmetry of the given theory~\cite{Sugano:1982bm, Sugano:1986xb, Sugano:1989rq, Sugano:1991ir, Sugano:1991kd, Sugano:1991ke}.
Thus, if a gauge symmetry, a closed PB algebra, is violated, $\#$ of first-class constraints reduces. 
It provides new second-class constraints and/or non-linear DOFs by the above definition. 
We remark that the non-linear DOFs provide the upper bound of the possible number of propagating DOFs. 
Inherited this property, the non-linear DOFs are sometimes called the full/total DOFs of a given theory. 
In the next subsection, we introduce NGR, in which the local LI violates~\cite{Tomonari:2024lpv, Tomonari:2024ybs}. 

For irregular systems, we need to regularize the system to count the DOFs, although there is no generic method. 
Here, we call a theory irregular if it has a constraint that violates the property of regularity.
Several examples can be verified in Refs.~\cite{Miskovic:2003ex, Tomonari:2024lpv}.
In general, the regularized system does not change only in a local region of the constraint surface.
That is, the method of regularization is different in each region of the constraint surface. 
This means that we cannot use the DOFs provided by the result of the DB analysis to set the upper bound of the perturbation theory.
In the following, we abbreviate ``w'' on top of each quantity such as the Weitzenb\"{o}ck connection, $\overset{\mathrm{w}}{\Gamma}{}^{\rho}_{\mu\nu}$, and the torsion tensor, $\overset{\mathrm{w}}{T}{}^{\rho}{}_{\mu\nu}$, for simplicity.

\subsection{\label{02:02}Violation of Local Lorentz invariance in NGR and Extra Degrees of Freedom}

The Lagrangian of NGR is given as follows~\cite{Hayashi:1979qx}:
\begin{equation}
\label{Lagrangian of NGR}
\begin{split}
    L_\mathrm{NGR}
    := 
    \theta^{-1}\,\mathcal{L}_\mathrm{NGR} 
    &= c_{1}\,T^{\mu\nu\rho}\,T_{\mu\nu\rho} + c_{2}\,T^{\mu\nu\rho}\,T_{\rho\mu\nu} + c_{3}\,T^{\mu}{}_{\mu\rho}\,T_{\nu}{}^{\nu\rho} \,\\
    &= c_{1}\,g_{\mu\sigma}\,g^{\nu\lambda}\,g^{\rho\kappa}\,T^{\mu}{}_{\lambda\kappa}\,T^{\sigma}{}_{\nu\rho} + c_{2}\,g^{\nu\lambda}\,T^{\mu}{}_{\lambda\rho}\,T^{\rho}{}_{\mu\nu} + c_{3}\,g^{\rho\nu}\,T^{\mu}{}_{\mu\rho}\,T^{\lambda}{}_{\lambda\nu} \,,
\end{split}
\end{equation}
where $\theta$ is the determinant of the co-vierbein field components and $c_{1}$, $c_{2}$, $c_{3}$ are three free parameters that range in real value. 
For example, in TEGR, which is equivalent to GR up to a boundary term in the action integral except for the geometry that describes spacetime, $c_{1}=-1/4$, $c_{2}=1/2$, $c_{3}=1$.

Applying the variational principle with respect to the (co-)vierbein field under the imposition of Dirichlet boundary conditions on these fields, we obtain the field equation
\begin{equation}
\begin{split} 
    \frac{1}{2}\,e_{A}{}^{\rho}\,\mathcal{T}_{\rho}{}^{\nu} 
    &=
    \theta^{-1}\,\partial_{\mu}(\theta\,e_{A}{}^{\rho}\,S_{\rho}{}^{\mu\nu}) 
    + e_{A}{}^{\lambda}\,T^{\rho}{}_{\mu\lambda}\,S_{\rho}{}^{\mu\nu} 
    - \frac{1}{2}\,e_{A}{}^{\nu}\,L_\mathrm{NGR} 
    \,,\\
    S_{\rho}{}^{\mu\nu} 
    &:= 
    c_{1}\,T_{\rho}{}^{\mu\nu} 
    + c_{2}\,T^{[\mu\nu]}{}_{\rho} 
    + c_{3}\,\delta_{\rho}{}^{[\mu}T^{|\kappa|}{}_{\kappa}{}^{\nu]} 
    \,.
\end{split}
\label{field equation of NGR}
\end{equation}
Here, $e_{A}{}^{\rho}\,\mathcal{T}_{\rho}{}^{\nu} = \theta^{-1}\,\delta\mathcal{L}_\mathrm{matter}\,/\,\delta\,e^{A}{}_{\nu}$ is the push forward of energy-momentum tensor by the vierbein field. 
Pulling back this equation from the internal space to spacetime and taking the anti-symmetric part of this field equation, we obtain 
\begin{equation}
    0 =
    \left( - 2\,c_{2} + c_{3} \right)\,\overset{\circ}{\nabla}{}_{\rho}T^{\rho}{}_{[\mu\nu]} 
    -\left( 2\,c_{1} + c_{2} \right)\,\overset{\circ}{\nabla}{}_{\rho}T_{[\mu\nu]}{}^{\rho} 
    + \left( c_{1} - \frac{1}{2}\,c_{2} + \frac{1}{2}\,c_{3}\right)\,T^{\rho\sigma}{}_{[\mu}T_{\nu]\rho\sigma} 
    \,,
\label{anti-symmetric part of field equations of NGR}
\end{equation}
where the circle `` $\circ$ '' on top of $\nabla$ denotes the covariant derivative with respect to the Levi-Civita connection.
We remark $\mathcal{T}_{[\mu\nu]} = 0\,$. 
The number of independent components of Eq.~\eqref{anti-symmetric part of field equations of NGR} is six at most, and this number coincides with that of the generator of $SO(1\,,3)$-symmetry. 
In the TEGR case, Eq.~\eqref{anti-symmetric part of field equations of NGR} is, on the one hand, automatically satisfied, subject to automatic vanishment of each coefficient in Eq.~\eqref{anti-symmetric part of field equations of NGR} under $c_{1}=-1/4$, $c_{2}=1/2$, $c_{3}=1$. 
On the other hand, if Eq.~\eqref{anti-symmetric part of field equations of NGR} is not identically satisfied, the equation suggests the existence of propagating modes generated by the violation of the local LI. 

In the Hamiltonian formulation of NGR, we can classify the theory into the nine independent types based on the $SO(3)$-irreducible decomposition of canonical momentum~\cite{Blixt:2018znp}. 
DB analysis on each type of NGR was performed in Refs.~\cite{Tomonari:2024ybs, Tomonari:2024lpv} while satisfying the diffeomorphism invariance (hypersurface deformation algebra in terms of PB algebra)~\cite{Dirac:1958sc} in all types.
The authors in Refs.~\cite{Tomonari:2024ybs, Tomonari:2024lpv} clarified the constraint structure of each type, which is the expression of the internal local symmetry of each type in terms of PB algebra.
The resulting PB algebra shows that all types do not have the local LI except for Type 6 (TEGR).
Based on this, the authors counted out the non-linear DOFs of each type.
We summarize the result in Table~\ref{Table:fullDoFofNGR}. 

All types violate the local LI and give rise to extra DOFs up to six, except for Type 6 (TEGR)~\cite{Blagojevic:2000pi, Blagojevic:2000qs, Maluf:2000ag, Ferraro:2016wht}. 
Since all types satisfy the diffeomorphism invariance, the extra DOFs in each type should be ascribed to the violation of the local LI. 
This perspective is essential since it indicates that the propagating modes in NGR should be described in terms of the anti-symmetric part of the vierbein perturbation. 
In the next section, we revisit a complete formalism for describing linear perturbations of NGR around the flat FLRW spacetime with local LI. 
\begin{table}[ht!]
    \centering
    \renewcommand{\arraystretch}{1.5}
    \begin{tabular}{ c || c | c | c}
	Theory & Conditions on parameter space $(c_{1}\,,c_{2}\,,c_{3})$ & Non-linear DOF & Regularity\\ \hline \hline
	Type 1 & arbitrary & 8 & $-$\\ \hline
	Type 2 & $2\,c_{1} - c_{2} + c_{3} = 0$ & 6 & $\checkmark$\\ \hline
	Type 3 & $2\,c_{1} + c_{2} = 0$ & 5 & $\checkmark$\\ \hline
	Type 4 & $2\,c_{1} - c_{2} = 0$ & 5 & $\times$\\ \hline
	Type 5 & $2\,c_{1} - c_{2} + 3\,c_{3} = 0$ & 7 & $\checkmark$\\ \hline
	Type 6 (TEGR) & 
    \begin{tabular}{c}
        $2\,c_{1} - c_{2} + c_{3} = 0$ \\ $\&$ $2\,c_{1} + c_{2} = 0$
    \end{tabular} 
    & 2 & $\checkmark$\\ \hline
	Type 7 & 
    \begin{tabular}{c}
        $2\,c_{1} + c_{2} = 0$ \\ 
        $\&$ $2\,c_{1} - c_{2} = 0$
    \end{tabular} 
    &  0 (Topological in bulk spacetime) & $\times$ \\ \hline
	Type 8 & 
    \begin{tabular}{c}
        $2\,c_{1} + c_{2} = 0$ \\ 
        $\&$ $2\,c_{1} - c_{2} + 3\,c_{3} = 0$
    \end{tabular}  
    & 6 (Generic) or 4 (Special) & $\checkmark$\\ \hline
	Type 9 & 
    \begin{tabular}{c}
        $2\,c_{1} - c_{2} + c_{3} = 0$ \\ 
        $\&$ $2\,c_{1} - c_{2} = 0$ \\ 
        $\&$ $2\,c_{1} - c_{2} + 3\,c_{3} = 0$ 
    \end{tabular} 
    & 3 & $\checkmark$\\ \hline
    \end{tabular}
    \caption{
    Conditions on parameters, $c_{1}$, $c_{2}$, $c_{3}$, and nonlinear DOFs of each type of NGR in the $SO(3)$-irreducible decomposition of canonical momentum. 
    We remark that the sign of the parameter $c_{2}$ is opposite to that of the original work~\cite{Blixt:2018znp}. 
    ``Special'' denotes the case that occurs only under the satisfaction of a set of specific conditions on Lagrange multipliers, whereas ``Generic'' denotes the case without any conditions. 
    ``Regularity'' means the closedness as a linear combination in the first-order variation of each constraint with respect to all constraints existing in a theory~\cite{Dirac:1950pj,Dirac:1958sc,Dirac:1958sq,Miskovic:2003ex}.
    For details, see Refs.~\cite{Tomonari:2024ybs, Tomonari:2024lpv}.
    }
    \label{Table:fullDoFofNGR}
\end{table}

\section{\label{03}Cosmological Perturbation from the viewpoint of Local Lorentz Invariance}

We revisit the conventional framework of linear perturbations to clarify how each perturbation field is related to local Lorentz invariance.
This step is important because, if one fixes a perturbation field associated with local Lorentz invariance in a theory where this invariance is violated, such gauge fixing may inadvertently remove a genuine propagating mode.

Cosmological perturbations are considered around the flat FLRW spacetime:
\begin{equation}
    ds^{2} = - dt^{2} + a^{2}\delta_{ij}dx^{i}dx^{j},
\label{flat FLRW-spacetime}
\end{equation}
where $a = a(t)$ is the scale factor, and $i$ and $j$ run from $1$ to $3$. 
In Refs.~\cite{Wu:2012hs,Chen:2010va}, the authors decompose the co-vierbein field, $e^{I}{}_{\mu}$, into the symmetric part, $\bar{e}{}^{I}{}_{\mu}$, and the anti-symmetric part, $\tilde{e}{}^{I}{}_{\mu}\,$.
The symmetric and the anti-symmetric parts describe the DOFs of the metric tensor and the extra DOFs generated by the violation of the local LI, respectively. 
That is, we can split $e^{I}{}_{\mu}$ into as follows:
\begin{equation}
    e^{I}{}_{\mu} = \bar{e}{}^{I}{}_{\mu} + \tilde{e}{}^{I}{}_{\mu}\,.
\label{vierbein decomposition}
\end{equation}
According to Refs.~\cite{Wu:2012hs, Chen:2010va}, we introduce perturbations in terms of the vierbein field with the condition: $g_{\mu\nu} = e^{I}{}_{\mu}e^{J}{}_{\nu}\eta_{IJ} = \bar{e}^{I}{}_{\mu}\bar{e}^{J}{}_{\nu}\eta_{IJ}$. Here, $\eta_{IJ} = \mathrm{diag}\,(-1, +1, +1, +1)$ is the Minkowski metric. 

In Ref.~\cite{Izumi:2012qj}, the authors indicated that the perturbation theory formulated in Refs.~\cite{Chen:2010va, Dent:2010nbw, Wu:2012hs} is incomplete since the anti-symmetric tensor introduced in Refs.~\cite{Chen:2010va, Dent:2010nbw, Wu:2012hs} can be further decomposed into a pseudo-scalar and a transverse pseudo-vector.
The authors also redefined all perturbed fields and derived the gauge transformation of these fields. 
However, the literature confuses the symmetric and the anti-symmetric part of the co-vierbein decomposition and imposes gauge conditions that fix a part of the anti-symmetric components of the co-vierbein. 
Our purpose is to investigate the propagation of the anti-symmetric parts of the co-vierbein, which represent the propagating modes generated by the violation of local LI.
This property indicates that we should not fix the gauge corresponding to the anti-symmetric part of the co-vierbein at least in advance. 
Thus, we must clarify the origin of each perturbation field in the pioneering work~\cite{Izumi:2012qj}. 

For our purpose, we reconsider the co-vierbein perturbation in Refs.~\cite{Wu:2012hs, Chen:2010va, Izumi:2012qj}. 
Let $\bar{e}{}^{I}{}_{\mu}$ and $\tilde{e}{}^{I}{}_{\mu}$ decompose as follows:
\begin{equation}
\begin{split}
    &\bar{e}{}^{0}{}_{\mu} 
    = 
    \left(1 + \psi\right)\,\delta^{0}{}_{\mu} 
    + a\,\left( \partial_{i}F + G_{i} \right)\,\delta^{i}{}_{\mu}
    \,,\\
    &\bar{e}{}^{a}{}_{\mu} 
    =
    a\,\left(1 - \varphi\right)\,\delta^{a}{}_{\mu} 
    + a\,\delta^{ak}\,\left(h_{jk} 
    + \partial_{j}\partial_{k}B + \partial_{j}C_{k} + \partial_{k}C_{j}\right)\,\delta^{j}{}_{\mu}
    \,,
\end{split}
\label{symmetric part of vierbein perturbations}
\end{equation}
and
\begin{equation}
\begin{split}
    &\tilde{e}{}^{0}{}_{\mu} 
    =
    a\,\left(\partial_{i}\alpha + \alpha_{i}\right)\,\delta^{i}{}_{\mu}
    \,,\\
    &\tilde{e}{}^{a}{}_{\mu} 
    = 
    a\,\delta^{ai}\,\delta^{kl}\,\epsilon_{ijk}\,
    \left(\partial_{l}\tilde{\sigma} + \tilde{V}_{l}\right)\,\delta^{j}{}_{\mu}
    \,.
\end{split}
\label{antisymmetric part of vierbein perturbations}
\end{equation}
$\psi$, $\varphi$, $B$, and $F$ are the scalar perturbations of the symmetric part of the co-vierbein field, $\bar{e}^{I}{}_{\mu}$.
$\alpha$ and $\tilde{\sigma}$ are the scalar and pseudo-scalar perturbations of the anti-symmetric part of the co-vierbein field, $\tilde{e}{}^{I}{}_{\mu}$.
$C_{i}$ and $G_{i}$ are the transverse vector perturbations of the symmetric part of the co-vierbein field, and $\alpha_{i}$ and $\tilde{V}_{i}$ are the transverse vector and the pseudo-vector perturbations of the anti-symmetric part of the co-vierbein field, respectively. 
$h_{ij}$ is the traceless and transverse spatial tensor perturbation, which corresponds to gravitational waves in GR. 
We note that the local Lorentz transformation acts on the first and second equations in Eq.~\eqref{antisymmetric part of vierbein perturbations} as a boost and a rotation, respectively.
Combining Eqs.~\eqref{symmetric part of vierbein perturbations} and \eqref{antisymmetric part of vierbein perturbations} based on Eq.~\eqref{vierbein decomposition}, we obtain
\begin{equation}
\begin{split}
    &e^{0}{}_{\mu} 
    = 
    \left(1 + \psi\right)\,\delta^{0}{}_{\mu} + a\,\left[ \partial_{i}(F + \alpha) + \left(G_{i} + \alpha_{i}\right) \right]\,\delta^{i}{}_{\mu} 
    \,,\\
    &e^{a}{}_{\mu} 
    = 
    a\,\left(1 - \varphi\right)\,\delta^{a}{}_{\mu} 
    + a\,\delta^{ai}\,
    \left[
        h_{ji} 
        + \partial_{j}\partial_{i}B 
        + \partial_{j}C_{i} 
        + \partial_{i}C_{j} 
        + \epsilon_{ijk}\,\delta^{kl}\,
        \left(\partial_{l}\tilde{\sigma} + \tilde{V}_{l}\right)
    \right]\,\delta^{j}{}_{\mu} 
    \,.
\end{split}
\label{wrong vierbein perturbations}
\end{equation}
We remark that we do not confuse the spatial indices of spacetime, $i\,,j\,,k\,,\cdots\,$, and the Lorentz indices, $a\,,b\,,c\,,\cdots\,$.

However, this decomposition is not well-defined since the functional DOF of each side of the equations in Eq.~\eqref{wrong vierbein perturbations} does not match one another.
That is, $e^{0}{}_{\mu}$ can encapsulate four DOFs, whereas the right-hand side has seven perturbation fields.
In the second equation of Eq.~\eqref{wrong vierbein perturbations}, the co-vierbein has room to encapsulate these exceeded perturbation fields.
Thus, to reconcile this inconsistency, we modify it as follows:
\begin{align}
\label{vierbein perturbations}
\begin{split}
    &e^{0}{}_{\mu} 
    = 
    \left(1 + \psi\right)\,\delta^{0}{}_{\mu} + a\,\left(\,\partial_{i}\alpha + \alpha_{i}\,\right)\,\delta^{i}{}_{\mu} 
    \,,\\
    &e^{a}{}_{\mu} 
    = 
    a\,\left(1 - \varphi\right)\,\delta^{a}{}_{\mu} 
    + \delta^{ai}\,(\partial_{i}F + G_{i})\,\delta^{0}{}_{\mu} 
    \\
    & \qquad \qquad
    + a\,\delta^{ai}\,
    \left[
        h_{ji} + \partial_{j}\partial_{i}B + \partial_{j}C_{i} + \partial_{i}C_{j} 
        + \epsilon_{ijk}\,\delta^{kl}\,\left(\partial_{l}\tilde{\sigma} + \tilde{V}_{l}\right)
    \right]
    \,\delta^{j}{}_{\mu} 
    \,.
\end{split} 
\end{align}
This perturbation theory can describe 10 (spacetime sector: $h_{ij}$, $\psi$, $\varphi$, $F$, $B$, $G_{i}$, $C_{i}$) + 6 (internal-space sector: $\alpha$, $\tilde{\sigma}$, $\alpha_{i}$, $\tilde{V}_{i}$) = 16 propagating DOFs at most. 
In particular, a violation of local LI causes possible propagating DOFs that are ascribed to the internal space. 
This decomposition is none other than that provided in Ref.~\cite{Izumi:2012qj} except for the difference in notation. 
Compared with Refs.~\cite{Golovnev:2018wbh, Golovnev:2023ddv, Golovnev:2023jnc}, our parameterization coincides with theirs, except for the use of conformal time, the non-symmetrization of $\partial_{i} C_{j}$ terms, and the composition of variables that the local Lorentz transformation operates on.\footnote{
Translating their notations into ours, it states that scalar $\alpha + F$, pseudo-scalar $\tilde{\sigma}\,$, vector $\alpha_{i} + G_{i}$, and pseudo-vector $\tilde{V}_{i}$ correspond to local Lorentz rotation of the vierbein in Ref.~\cite{Golovnev:2018wbh}.
Our current consideration does not follow from this reference for the following three reasons.
a) $\alpha$, $F$, $\tilde{\sigma}$, $\alpha_{i}$, $G_{i}$, $\tilde{V}_{i}$ should be counted separately.
Otherwise, the functional degrees of freedom of the vierbein components and the total number of perturbation fields do not match each other, implying that we implicitly induce constraints.
This may accidentally change the original theory into a different one.
b) local LI acts on $\tilde{V}_{i} + \partial_{i}\tilde{\sigma}$ for three rotations, but so does not for $\alpha_{i} + G_{i} + \partial_{i}(\alpha + F)$ for three boosts due to the incorporation of $G_{i}$ and $F$.
c) $G_{i}$ and $F$ correspond to the components of the metric perturbation, indicating that these fields do not contribute to internal-space symmetries.
}

Now, we consider the gauge transformation of these perturbed fields.
For infinitesimal small coordinate transformations, $x^{\mu} \rightarrow x'^{\mu} = x^{\mu} + \xi^{\mu}(x)\,$, the variation of the co-vierbein is given as follows:
\begin{equation}
\label{gauge transformation of perturbations}
    \delta_{\xi}\,e^{I}{}_{\mu} 
    = 
    - \mathcal{L}_{\xi}\,e^{I}{}_{\mu} 
    = 
    -\xi^{\nu}\partial_{\nu}e^{I}{}_{\mu} -e^{I}{}_{\nu}\,\partial_{\mu}\xi^{\nu}
    \, .
\end{equation}
Here, we denote by $\mathcal{L}_{X}Y$ the Lie derivative of $Y$ with respect to $X$.
Then, the gauge transformations of the perturbations are calculated as follows:
\begin{equation}
    e^{I}{}_{\mu} \rightarrow e'^{I}{}_{\mu} = e^{I}{}_{\mu} + \delta_{\xi}e^{I}{}_{\mu} = e^{I}{}_{\mu} -\xi^{\nu}\partial_{\nu}e^{I}{}_{\mu} -e^{I}{}_{\nu}\,\partial_{\mu}\xi^{\nu}\,.
\label{}
\end{equation}
Decomposing the spatial component of $\xi^{\mu}$, $\xi^{i}$, further into $\xi^{i} = \partial^{i}\xi + \xi^{(v)}{}^{i}\,$, where $\xi^{(v)}{}^{i}$ is a transverse vector, we can derive following transformation rules
\begin{equation}
\begin{split}
    &\psi' 
    = 
    \psi - \dot{\xi}{}^{0}
    \,,\\
    &\varphi' 
    =
    \varphi + \frac{\dot{a}}{a}\,\xi^{0} 
    \,,\\
    &\alpha' 
    =
    \alpha - \frac{1}{a}\xi^{0}
    \,,\\
    &B' 
    = 
    B - \frac{1}{a}\xi
    \,,\\
    &F' 
    = 
    F - \left(\dot{\xi} - \frac{\dot{a}}{a}\xi\right)
    \,,\\
    &\tilde{V}'{}_{i} 
    =
    \tilde{V}{}_{i} 
    - \frac{1}{a}\epsilon_{ijk}\delta^{jl}\delta^{km}\partial_{l}\xi^{(v)}_{m}
    \,,\\
    &C'_{i} 
    = 
    C_{i} - \frac{1}{2a}\xi^{(v)}_{i}
    \,, \\
    &G'_{i} 
    =
    G_{i} - \left( \dot{\xi}{}^{(v)}_{i} - \frac{\dot{a}}{a}\,\xi^{(v)}_{i}\right)
    \,.
\end{split}
\label{gauge transformation of perturbations in explicit form}
\end{equation}
Here, the dot `` $\dot{}$ '' stands for the time derivative.
Other perturbations, $h_{ij}$, $\alpha_{i}$, and $\tilde{\sigma}$ do not change with respect to the infinitesimal gauge transformation given by Eq.~\eqref{gauge transformation of perturbations}. 
This transformation coincides with that provided in the pioneering work~\cite{Izumi:2012qj} up to the difference in notation.\footnote{
In our notation, the sign of $\varphi$ is opposite. 
}
Based on the above results, we can compose five set of gauge-invariant variables as follows:
\begin{equation}
    \beta = \psi + \varphi - a \dot{\alpha}\,,
    \qquad 
    \gamma = \varphi + \dot{a}\alpha
    \qquad 
    A = F - a \dot{B}\,,
    \qquad 
    B_{i} = G_{i} - 2 a \dot{C}_{i}\,,
    \qquad
    \tilde{W}_{i} = \tilde{V}_{i} - 2 \epsilon_{i j k} \partial_{j}{C_{k}}\,.
\label{gauge invarinant variables}
\end{equation}
Taking $h_{ij}$, $\alpha_{i}$, and $\tilde{\sigma}$ into account, our theory consists of twelve gauge-invariant variables in total; we can replace twelve out of sixteen perturbation variables in terms of these gauge-invariant variables. 

Now, the origin of each perturbation field and its role are clear. 
We shall consider gauge-fixing conditions for our analysis. 
Fixing the gauge $\xi^{0}$, one of the three scalar perturbations, $\psi$, $\varphi$, or $\alpha$, vanishes from the theory. 
For the gauge $\xi$, one of the two scalar perturbations, either $B$ or $F$, vanishes. 
In the same way, fixing the gauge $\xi^{(v)}{}_{i}$, one of the pseudo-vector $\tilde{V}_{i}$, the vector $C_{i}$, or the vector $G_{i}$ vanishes.
If a given theory satisfies both diffeomorphism and local LI, we can simplify the perturbation theory as desired by fixing the gauge freely. 
In NGR, however, local LI is violated while preserving diffeomorphism invariance. 
Therefore, we should not fix the following perturbations that originated from the violation of local LI: $\alpha$, $\alpha_{i}$, $\tilde{\sigma}$, and $\tilde{V}_{i}$.  

To formulate a perturbation theory for investigating the propagating DOFs in NGR, we shall consider an appropriate gauge choice. 
For the gauge $\xi^{0}$, we should fix it so that either of $\psi = 0$ or $\varphi = 0$ holds. 
In our work, we choose the gauge to realize $\varphi = 0$. 
For the gauges $\xi$ and $\xi^{(v)}{}_{i}$, in this work, we fix them for $B = 0$ and $C_{i} = 0$, respectively.
Therefore, the possible propagating DOFs are $h_{ij}$, $\psi$, $F$, and $G_{i}$ of the symmetric part of the co-vierbein field and $\alpha$, $\alpha_{i}$, $\tilde{\sigma}$, and $\tilde{V}_{i}$ of the anti-symetric part of the co-vierbein field. 
Explicitly, the co-vierbein field turns into 
\begin{equation}
\begin{split}
    &e^{0}{}_{0} 
    = 
    1 + \psi \,,\\
    &e^{0}{}_{i} 
    = 
    a\left( \partial_{i}\alpha + \alpha_{i}\right) 
    \,,\\
    &e^{a}{}_{0} 
    = 
    \delta^{ai}\,\left( \partial_{i}F + G_{i} \right) 
    \,,\\
    &e^{a}{}_{i} 
    = 
    a\,\delta^{a}{}_{i} + a\,\delta^{al}\,
    \left[
        h_{li} + \epsilon_{lij}\,\delta^{jk}\,\left(\partial_{k}\tilde{\sigma} + \tilde{V}_{k}\right)
    \right]
    \, .
\end{split}
\label{gauge fixed vierbein perturbations}
\end{equation}
The above gauge choice is well-known as the spatially flat gauge.

Finally, we note the following three points. 
1) The authors in Ref.~\cite{Izumi:2012qj} indicates that in a parity-preserving theory the possible coupling of the pseudo-vector perturbation, $\tilde{V}_{i}$, with the vector perturbation, $\alpha_{i}$, is $\epsilon_{ijk}(\partial_{i}\alpha_{j})\tilde{V}_{k}$ only.
Here, we modified their notation to ours.
2) In the linear perturbation theory, we can treat all modes separately except for the vector and pseudo-vector perturbations.\footnote{
Technically, the term $\partial_{i}{\tilde{\sigma}}$ behaves as a pseudo-vector perturbation, suggesting that $\tilde{\sigma}$ is independent of any of the other scalar perturbations in linear perturbations. 
}
3) The background Lagrangian density of NGR becomes
\begin{equation}
\label{background Lagrangian}
    \mathcal{L}^\mathrm{(flat\,FLRW)}_\mathrm{NGR} 
    = 
    - 3\,( 2c_{1} - c_{2} + 3 c_{3} )\,a^{3} H^{2}
\end{equation}
where $H := \dot{a}/a$ is Hubble parameter. 
In particular, for Types 5, 8, and 9, the background Lagrangian density vanishes; that is, the existence of matter contributes to the time evolution of spacetime more than first order.
We note that Types 2 and 3 are none other than Type 6 (TEGR) with a violation of local LI at least partially (for details, see Ref.~\cite{Tomonari:2024ybs}.
In the following sections, we utilize this decomposition of the co-vierbein field, Eq.~\eqref{gauge fixed vierbein perturbations}, to investigate propagating DOFs of each type of NGR up to the second-order perturbations.

\section{\label{04}Background Equation with Matter Field}

We introduce a scalar field, $\Phi\,$, as matter source:
\begin{equation}
    L_\mathrm{matter} = \theta^{-1}\,\mathcal{L}_\mathrm{matter} 
    = 
    - \frac{1}{2}\,g^{\mu\nu}\,\partial_{\mu}{\Phi}\,\partial_{\nu}{\Phi} - V(\Phi) \,.
\label{massless test matter density}
\end{equation}
Here, $\theta$ is the determinant of the co-frame field. 
Varying with respect to $\Phi$, we of course obtain the field equation:
\begin{equation}
    \theta^{-1}\partial_{\mu}(\theta g^{\mu\nu}\partial_{\nu}\Phi) - V' = 0 \,.
\label{field equ of Phi}
\end{equation}
According to the pioneering work~\cite{Izumi:2012qj}, we decompose $\Phi$ into the background and the first-order perturbation part, $\Phi_{0}$ and $\delta\Phi$, respectively, as follows:
\begin{equation}
    \Phi = \Phi_{0} + \delta\Phi \,.
\label{massless test matter perturbation}
\end{equation}
We can expand the potential term, $V$, up to second order with respect to $\delta\Phi$ as follows:
\begin{equation}
    V(\Phi_{0} + \delta\Phi) = V_{0} + V'_{0}\,\delta\Phi + \frac{1}{2}\,V''_{0}\,(\delta\Phi)^{2} \,,
\label{expansion of V up to 2nd order}
\end{equation}
where the prime $'$ stands for a derivative with respect to $\Phi\,$. 
We express $V_{0} := V(\Phi_{0})$, $V'_{0} = V'(\Phi_{0})$, and $V''_{0} = V''(\Phi_{0})$ for simplicity.
The isometry of the background flat-FLRW spacetime requires that the spatial derivative of the background scalar field $\Phi_0$ vanishes: 
\begin{equation}
    \partial_{i}{\Phi_{0}} = 0 \,.
\label{matter with isommetry}
\end{equation}

Based on this set-up, the field equations of NGR in Eq.~\eqref{field equation of NGR} around the flat FLRW background spacetime are given as follows:
\begin{equation}
\begin{split}
    & -3 \left( 2 c_{1} - c_{2} + 3 c_{3} \right) H^{2} 
    = 
    - \frac{1}{2} \dot{\Phi}_{0}\dot{\Phi}_{0} + V_{0}
    \,, \\
    & - (2 c_{1} - c_{2} + 3 c_{3}) \left( \frac{3}{2} H^{2} + \dot{H} \right) 
    = 
    -\frac{1}{2} V_{0} + \frac{1}{4} \dot{\Phi}_{0} \dot{\Phi}_{0}
    \,.
\end{split}
\label{simplified background equation of NGR with non-zero dotPhi}
\end{equation}
We can verify that these equations coincide with those in GR by setting $c_{1} = - 1/4$, $c_{2} = 1/2$, $c_{3} = 1$. 
Taking into account the result of its DB analysis~\cite{Tomonari:2024lpv, Tomonari:2024ybs}, the violation of local LI appears in the weight, $2 c_{1} - c_{2} + 3 c_{3}\,$, which couples matters with gravitation. 
In addition, combining them, we obtain an equation to describe the time evolution of the Hubble parameter:
\begin{equation}
    - (2 c_{1} - c_{2} + 3 c_{3}) \dot{H}  
    =
    \frac{1}{2} \dot{\Phi}_{0} \dot{\Phi}_{0} - V_{0} \,.
\label{simplified background equation of NGR wrt SS with non-zero dotPhi}
\end{equation}
If $2 c_{1} - c_{2} + 3 c_{3} = 0$ holds, that is, in Types 5, 8, and 9, the left-hand sides of two equations in Eq.~\eqref{simplified background equation of NGR with non-zero dotPhi} vanish. 
Therefore, the Hubble parameter is arbitrary, and in Types 5, 8, and 9, we cannot specify the background spacetime.
This result can be interpreted as indicating that matter does not contribute to the time evolution of the background spacetime.
Therefore, perturbative analyses in these types effectively correspond to studying the propagating degrees of freedom just on maximally symmetric spacetime;  
these models are not suitable for cosmological applications.
Nevertheless, the possibility of astrophysical applications, such as to black hole spacetime, would still remains open. 

Finally, the background field equation of the matter field, Eq.~\eqref{field equ of Phi}, around the background spacetime becomes 
\begin{equation}
    - \ddot{\Phi}_{0} - 3 H \dot{\Phi}_{0} - V'_{0} = 0 \,,
\label{background equation of test matter}
\end{equation}
We use these equations, Eqs.~\eqref{simplified background equation of NGR with non-zero dotPhi} and~\eqref{background equation of test matter}, to eliminate the background variable, $\Phi_{0}$, from the perturbed Lagrangian density in the subsequent sections.
As a note, combining Eq.~\eqref{simplified background equation of NGR wrt SS with non-zero dotPhi} with Eq.~\eqref{background equation of test matter}, we can solve the Hubble parameter and the background matter field. 
Thus, if a second-order perturbed Lagrangian density contains a term that consists only of the background matter field and the Hubble parameter, or equivalently, the scale factor, we can freely drop it without loss of generality under the satisfaction of the background equations. 
We also use this property in the subsequent sections.

\section{\label{05}Propagating Modes}

\subsection{\label{05:01}Tensor Perturbation}

We calculate the tensor perturbation of NGR up to second order. 
The Lagrangian of the theory is given by Eq.~\eqref{Lagrangian of NGR}. 
Focusing on the tensor terms of Eq.~\eqref{gauge fixed vierbein perturbations}, we find that the co-vierbein and vierbein field components are derived as follows~\cite{Izumi:2012qj}:
\begin{align}
\label{vierbein in TPs}
\begin{split}
    &e^{0}{}_{\mu}\,dx^{\mu} = dt
    \,,\\
    &e^{a}{}_{\mu}\,dx^{\mu} = a\,\delta^{ai}\,\left(\delta_{ij} + h_{ij}\right)dx^{j} 
    \,,\\
    &e_{0}{}^{\mu}\,\frac{\partial}{\partial x^{\mu}} = \frac{\partial}{\partial \,t} 
    \,,\\
    &e_{a}{}^{\mu}\,\frac{\partial}{\partial x^{\mu}} = a^{-1}\,\delta_{ai}\,\left(\delta^{ij} - h^{ij} + \delta_{kl}\,h^{ik}h^{jl}\right)\,\frac{\partial}{\partial\,x^{j}} 
    \,.
\end{split}
\end{align}
Using these formulae, the metric and its inverse tensor components are calculated as follows:
\begin{align}
\label{metric in TPs}
\begin{split}
    &g_{\mu\nu}\,dx^{\mu}dx^{\nu} 
    = 
    -dt^{2} + a^{2}\,\left(\delta_{ij} + 2\,h_{ij} + \delta^{kl}\,h_{ik}\,h_{jl}\right)\,dx^{i}dx^{j} 
    \,,\\
    &g^{\mu\nu}\,\frac{\partial}{\partial\,x^{\mu}}\frac{\partial}{\partial\,x^{\nu}} 
    =
    -\,\frac{\partial}{\partial\,t}\frac{\partial}{\partial\,t} + a^{-2}\,\left(\delta^{ij} -2\,h^{ij} + 3\,\delta_{kl}\,h^{ik}\,h^{jl}\right)\,\frac{\partial}{\partial\,x^{i}}\,\frac{\partial}{\partial\,x^{j}} 
    \,,
\end{split}
\end{align}
where we used the relations: $g_{\mu\nu} = e^{I}{}_{\mu}\,e^{J}{}_{\nu}\,\eta_{IJ}$ and $g^{\mu\nu} = e_{I}{}^{\mu}\,e_{J}{}^{\nu}\,\eta^{IJ}\,$. 
The determinant of the co-vierbein field components with the traceless gauge is\footnote{
We used the following formula:
\begin{align}
\label{formula for expanding the det of metric tensor}
    \det(1 + \epsilon A)
    = 
    \sum_{k=0}^{\infty}\frac{1}{k!}
    \left( - \sum_{j=1}^{\infty}\,\epsilon^{j}\,\frac{(-1)^{j}}{j}\,\mathrm{tr}(A^{j}) \right)^{k} 
    = 
    1 + \epsilon\,\mathrm{tr}(A) 
    + \frac{1}{2}\,\epsilon^{2}\,
    \left[\,\mathrm{tr}(A)^{2} - \,\mathrm{tr}(A^{2})\right] 
    + O(\epsilon^{3}) 
    \,,
\end{align}
where $\epsilon$ is an infinitesimal parameter.
}
\begin{equation}
    \theta = a^{3}\,\left(1 - \frac{1}{2}\,\delta^{ik}\,\delta^{jl}\,h_{ij}\,h_{kl}\right) \,.
\label{determinant of co-vierbein in TPs}
\end{equation}
The tensor perturbation of Eq.~\eqref{Lagrangian of NGR} up to second order becomes as follows:
\begin{align}
\label{TPs of NGR}
\begin{split}
    \mathcal{L}^\mathrm{(TP)}_\mathrm{NGR} 
    &=
    - 3 \left(2 c_{1} - c_{2} + 3 c_{3}\right) a^{3} H^{2}
    \\
    &\qquad 
    - a^{3}\,\left(2\,c_{1} - c_{2}\right)\,\delta^{ik}\,\delta^{jl}\,\dot{h}_{ij}\,\dot{h}_{kl} 
    + 2\,a^{3}\,H\,\left(2\,c_{1} - c_{2} + 3\,c_{3}\right) \,\delta^{ik}\,\delta^{jl}\,\dot{h}_{ij}\,h_{kl} 
    \\
    &\qquad 
    - \frac{1}{2}\,a^{3}\,H^{2}\,\delta^{ik}\,\delta^{jl}\,h_{ij}\,h_{kl} 
    + a\,\left(2\,c_{1} + c_{2}\right)\,\delta^{il} \,\delta^{jm} \,\delta^{kn} \,\partial_{i}h_{jk}\,\partial_{l}h_{mn} 
    \\
    &\qquad 
    - a\,\left(2\,c_{1} + c_{2}\right) \,\delta^{im} \,\delta^{jn} \,\delta^{kl} \,\partial_{i}h_{jk}\,\partial_{l}h_{mn}
    \,,
\end{split}
\end{align}
We used the torsion tensor components given in Appendix~\ref{app:01} and applied the transverse and traceless gauge.
In TEGR, where the parameters are chosen as $c_{1} = -1/4$, $c_{2} = 1/2$, $c_{3} = 1$, the overall sign of the kinetic term in Eq.~\eqref{TPs of NGR} is $-\,(2\,c_{1} - c_{2}) = +\,1\,$. 
From Table~\ref{Table:fullDoFofNGR}, Types 4, 7, and 9 do not contain the kinetic term of the tensor mode, suggesting that these types are not simple extensions of GR.
This classification coincides with that of the recent work Ref.~\cite{Bahamonde:2024zkb}. 

Substituting Eqs.~\eqref{massless test matter perturbation}, \eqref{expansion of V up to 2nd order}, \eqref{metric in TPs}, and \eqref{determinant of co-vierbein in TPs} into Eq.~\eqref{massless test matter density}, we rewrite the matter Lagrangian density as follows:
\begin{equation}
\begin{split}
    \mathcal{L}^\mathrm{(TP)}_\mathrm{matter} 
    &= 
    \frac{1}{2} a^{3} \dot{\Phi}_{0} \dot{\Phi}_{0} 
    - a^{3} V_{0} 
    + a^{3} \dot{\Phi}_{0} \dot{\delta\Phi} 
    - a^{3} V'_{0} \delta\Phi 
    + \frac{1}{2} a^{3} \dot{\delta\Phi} \dot{\delta\Phi} 
    - \frac{1}{2} a \delta^{i j} \partial_{i}{\delta\Phi} \partial_{j}{\delta\Phi} 
    - \frac{1}{2} a^{3} V''_{0} \delta\Phi \delta\Phi 
\end{split}
\label{test matter in TPs}
\end{equation}
up to second order in terms of the perturbation fields. 
Combining the perturbed matter Lagrangian density with Eq.~\eqref{TPs of NGR} and applying the background equations Eqs.~\eqref{simplified background equation of NGR with non-zero dotPhi} and~\eqref{background equation of test matter}, we obtain the total Lagrangian density as follows:
\begin{equation}
\begin{split}
    & \mathcal{L}^\mathrm{(TP\,;\,2nd\,order)}_\mathrm{total} 
    = 
    \mathcal{L}^\mathrm{(TP)}_\mathrm{NGR} 
    + \mathcal{L}^\mathrm{(TP)}_\mathrm{matter} \\
    &= 
    - a^{3}\,\left(2\,c_{1}- c_{2}\right)\,\delta^{ik}\,\delta^{jl}\,\dot{h}_{ij}\,\dot{h}_{kl} 
    + 2 \left(2\,c_{1} - c_{2} + 3\,c_{3}\right)\,a^{3}\,H\,\delta^{ik}\,\delta^{jl}\,\dot{h}_{ij}\,h_{kl} \\
    &\qquad 
    - \frac{1}{2}\,a^{3}\,H^{2}\,\delta^{ik}\,\delta^{jl}\,h_{ij}\,h_{kl} 
    + a\,\left(2\,c_{1} + c_{2}\right)\,
    \delta^{il}\,\delta^{jm}\,\delta^{kn}\,\partial_{i}h_{jk}\,\partial_{l}h_{mn} 
    - a\,\left(2\,c_{1} + c_{2}\right) \,\delta^{im}\,\delta^{jn}\,\delta^{kl}\,\partial_{i}h_{jk}\,\partial_{l}h_{mn} 
    \\
    &\qquad 
    + \frac{1}{2} a^{3} \dot{\delta\Phi} \dot{\delta\Phi} - \frac{1}{2} a \delta^{i j} \partial_{i}{\delta\Phi} \partial_{j}{\delta\Phi} 
    - \frac{1}{2} a^{3} V''_{0} \delta\Phi \delta\Phi 
    \,,
\end{split}
\label{TPs of NGR in 2nd-order}
\end{equation} 
where we dropped the surface terms. 
If $2c_{1}- c_{2} \neq 0$, the first term in Eq.~\eqref{TPs of NGR in 2nd-order} remains.
As a result, we conclude that Types 1, 2, 3, 5, 6 (TEGR), and 8 contain the propagating tensor mode, although Types 4, 7, and 9 do not. 
Moreover, the coefficient of the kinetic term of the tensor mode gives us a ghost-free condition, 
\begin{equation}
    2 c_{1} - c_{2} < 0\,,
\label{ghost-free condition of tensor mode}
\end{equation}
and $c_{3}$ is arbitrary.

\subsection{\label{05:02}Scalar Perturbation}

We calculate the scalar perturbation of NGR up to second order. 
Focusing on the scalar terms of Eq.~\eqref{gauge fixed vierbein perturbations}, the co-vierbein, the vierbein, the metric, the inverse metric components, and the determinant of the co-vierbein are derived as follows:
\begin{equation}
\begin{split}
    e^{0}{}_{\mu}\,dx^{\mu} 
    &= 
    (1 + \psi) dt + a\,\partial_{i}{\alpha}\,dx^{i} 
    \,,\\
    e^{a}{}_{\mu}\,dx^{\mu} 
    &=
    \delta^{a i}\,\partial_{i}{F}\,dt + a\,\delta^{a j}\,\delta_{i j}\,dx^{i} 
    \,,\\
    e_{0}{}^{\mu}\,\frac{\partial}{\partial x^{\mu}} 
    &=
    \left[ 1 - \psi + \psi^{2} + \delta^{i j} \partial_{i}{F} \partial_{j}{\alpha} \right]\,\frac{\partial }{\partial t} + a^{-1}\,\left[ (1 - \psi)\,\delta^{i j}\,\partial_{j}{F} \right]\,\frac{\partial }{\partial x^{i}} 
    \,,\\
    e_{a}{}^{\mu}\,\frac{\partial}{\partial x^{\mu}} 
    &= 
    \left[ - (1 - \psi)\,\delta_{a}{}^{i}\,\partial_{i}{\alpha} \right]\,\frac{\partial }{\partial t} + a^{-1}\,\left[ \delta_{a}{}^{i} + \delta_{a}{}^{j}\,\partial_{j}{\alpha}\,\delta^{i k}\partial_{k}{F} \right]\,\frac{\partial }{\partial x^{i}} 
    \,,\\
    g_{\mu\nu}\,dx^{\mu}dx^{\nu} 
    &= 
    \left[ 
        - ( 1 + 2\,\psi + \psi^{2} ) 
        + \delta^{i j}\,\partial_{i}{F}\,\partial_{j}{F} 
    \right]\,dtdt 
    \\ 
    & \qquad
    + 2a\,
    \left[ 
        \partial_{i}{F} - (1 + \psi)\,\partial_{i}{\alpha} 
    \right]\,
    dtdx^{i} 
    + a^{2}\,
    \left[ 
        \delta_{i j} - \partial_{i}{\alpha}\,\partial_{j}{\alpha} 
    \right]\,
    dx^{i}dx^{j} 
    \,,\\
    g^{\mu\nu}\,\frac{\partial}{\partial x^{\mu}}\frac{\partial}{\partial x^{\nu}} 
    &= 
    \left[ 
        - ( 1 - 2\,\psi + 3\,\psi^{2} ) 
        - 2\,\delta^{i j}\,\partial_{i}{F}\,\partial_{j}{\alpha} 
        + \delta^{i j}\,\partial_{i}{\alpha}\,\partial_{j}{\alpha} 
    \right]\,
    \frac{\partial}{\partial t}\frac{\partial}{\partial t} 
    \\ 
    & \qquad
    + 2\,a^{-1}\,
    \left[ 
        (1 - \psi)\,\delta^{i j}\,\partial_{j}{F} 
        - (1 - 2\,\psi)\,\delta^{i j}\,\partial_{j}{\alpha} 
    \right]\,
    \frac{\partial}{\partial t}\frac{\partial}{\partial x^{i}} 
    \\ 
    & \qquad
    + a^{-2}\,
    \left[  
        \delta^{i j} 
        + \delta^{i k}\,\delta^{j l}\,\partial_{k}{F} \partial_{l}{F} 
        + 2\,\delta^{ik}\partial_{k}{F}\,\delta^{jl}\partial_{l}{\alpha} 
    \right]\,
    \frac{\partial}{\partial x^{i}}\frac{\partial}{\partial x^{j}} 
    \,,\\
    &\theta 
    = 
    a^{3} ( 1 + \psi - \delta^{i j} \partial_{i}{F}\partial_{j}{\alpha} ) 
    \,. 
\end{split}
\label{vierbein and metric in SPs}
\end{equation}
The torsion tensor components given in Appendix~\ref{app:02}

Differing from the tensor perturbation, we must apply not only the background equations but also constraints to decouple propagating modes. 
To show the latter usage, we display an intermediate step. 
After deriving the perturbed Lagrangian and adopting the background equations, we obtain 

\begin{equation}
\begin{split}
    & \mathcal{L}^\mathrm{(SP\,;\,2nd\,order)}_\mathrm{total} 
    = 
    \mathcal{L}^\mathrm{(SP)}_\mathrm{NGR} + \mathcal{L}^\mathrm{(SP)}_\mathrm{matter} 
    \\
    &= ( 2 c_{1} - c_{2} + c_{3} ) a^{3} \delta^{ij} \partial_{i}{\dot{\alpha}} \partial_{j}{\dot{\alpha}}
    + 4 ( 2 c_{1} - c_{2} + 3 c_{3} ) a^{3} H \delta^{ij} \partial_{i}{\dot{\alpha}} \partial_{j}{\alpha}
    - 3 ( c_{2} - 3 c_{3} ) a^{3} H^{2} \delta^{ij} \partial_{i}{\alpha} \partial_{j}{\alpha}
    \\
    &\qquad
    + \frac{1}{2} a^{3} \dot{\Phi}_{0} \dot{\Phi}_{0} \delta^{ij} \partial_{i}{\alpha} \partial_{j}{\alpha}
    + 2 a^{2} \dot{\Phi}_{0} \delta^{ij} \partial_{i}{\alpha} \partial_{j}{\delta\Phi}
    \\
    &\qquad
    + \psi \Bigg{[} 
    2 ( 2 c_{1} - c_{2} + c_{3} ) a^{2} \delta^{ij} \partial_{i}\partial_{j}{\dot{\alpha}}
    + 4 ( 2 c_{1} - c_{2} + 3 c_{3} ) a^{2} H \delta^{ij} \partial_{i}\partial_{j}{\alpha}
    - 6 ( 2 c_{1} - c_{2} + 3 c_{3} ) a^{2} \delta^{ij} \partial_{i}\partial_{j}{F}
    \\
    &\qquad \qquad \quad
    - 2 ( 2 c_{1} - c_{2} + c_{3} ) a^{2} H \delta^{ij} \partial_{i}\partial_{j}{F}
    - a^{3} \dot{\Phi}_{0} \dot{\delta\Phi}
    - a^{3} V'_{0} \delta\Phi
    \Bigg{]}
    \\
    &\qquad
    + \psi \Bigg{[}
    - ( 2 c_{1} - c_{2} + c_{3} ) a \delta^{ij} \partial_{i}\partial_{j}{\psi}
    - 3 ( 2 c_{1} - c_{2} + 3 c_{3} ) a^{3} H^{2} \psi
    + \frac{1}{2} a^{3} \dot{\Phi}_{0} \dot{\Phi}_{0} \psi
    \Bigg{]}
    \\
    &\qquad
    + F \Bigg{[}
    6 ( 2 c_{1} - c_{2} + 3 c_{3} ) a^{3} \delta^{ij} \partial_{i}\partial_{j}{\dot{\alpha}}
    + ( 2 c_{1} - 7 c_{2} + 21 c_{3} ) a^{3} H^{2} \delta^{ij} \partial_{i}\partial_{j}{\alpha}
    + 2 a^{2} \dot{\Phi}_{0} \delta^{ij} \partial_{i}\partial_{j}{\delta\Phi}
    - V_{0} a^{3} \delta^{ij} \partial_{i}\partial_{j}{\alpha}
    \\
    &\qquad \qquad \quad
    - \frac{1}{2} a^{3} \dot{\Phi}_{0} \dot{\Phi}_{0} \partial_{i}\partial_{j}{\alpha}
    \Bigg{]}
    \\
    &\qquad
    + F \Bigg{[}
    10 c_{1} a^{3} H^{2} \delta^{ij} \partial_{i}\partial_{j}{F}
    - ( 2 c_{1} - c_{2} + 2 c_{3} ) a \delta^{ij} \delta^{kl} \partial_{i}\partial_{j}\partial_{k}\partial_{l}{F}
    \Bigg{]}
    \\
    &\qquad
    + \frac{1}{2} a^{3} \dot{\delta\Phi} \dot{\delta\Phi}
    - \frac{1}{2} a \delta^{ij} \partial_{i}{\delta\Phi} \partial_{j}{\delta\Phi}
    - \frac{1}{2} a^{3} V''_{0} \delta\Phi \delta\Phi
    \,,
\end{split}
\label{2nd order SPs of NGR}
\end{equation}
where we integrated by parts and neglected surface terms. 

Varying with respect to $\psi$ and $F$, we derive the following constraint equations:
\begin{equation}
\begin{split}
    &0 = 2 ( 2 c_{1} - c_{2} + c_{3} ) a^{2} \delta^{ij} \partial_{i}\partial_{j}{\dot{\alpha}}
    + 4 ( 2 c_{1} - c_{2} + 3 c_{3} ) a^{2} H \delta^{ij} \partial_{i}\partial_{j}{\alpha}
    - 6 ( 2 c_{1} - c_{2} + 3 c_{3} ) a^{2} \delta^{ij} \partial_{i}\partial_{j}{F}
    \\
    &\qquad
    - 2 ( 2 c_{1} - c_{2} + c_{3} ) a^{2} H \delta^{ij} \partial_{i}\partial_{j}{F}
    - a^{3} \dot{\Phi}_{0} \dot{\delta\Phi}
    - a^{3} V'_{0} \delta\Phi
    \\
    &\qquad
    - 2 ( 2 c_{1} - c_{2} + c_{3} ) a \delta^{ij} \partial_{i}\partial_{j}{\psi}
    - 6 ( 2 c_{1} - c_{2} + 3 c_{3} ) a^{3} H^{2} \psi
    + a^{3} \dot{\Phi}_{0} \dot{\Phi}_{0} \psi
\end{split}
\label{constraint wrt psi of SPs}
\end{equation}
and
\begin{equation}
\begin{split}
    &0 = 6 ( 2 c_{1} - c_{2} + 3 c_{3} ) a^{3} \delta^{ij} \partial_{i}\partial_{j}{\dot{\alpha}}
    + ( 2 c_{1} - 7 c_{2} + 21 c_{3} ) a^{3} H^{2} \delta^{ij} \partial_{i}\partial_{j}{\alpha}
    + 2 a^{2} \dot{\Phi}_{0} \delta^{ij} \partial_{i}\partial_{j}{\delta\Phi}
    - V_{0} a^{3} \delta^{ij} \partial_{i}\partial_{j}{\alpha}
    \\
    &\qquad
    - \frac{1}{2} a^{3} \dot{\Phi}_{0} \dot{\Phi}_{0} \partial_{i}\partial_{j}{\alpha}
    + 20 c_{1} a^{3} H^{2} \delta^{ij} \partial_{i}\partial_{j}{F}
    - 2 ( 2 c_{1} - c_{2} + 2 c_{3} ) a \delta^{ij} \delta^{kl} \partial_{i}\partial_{j}\partial_{k}\partial_{l}{F}\,.
\end{split}
\label{constraint wrt F of SPs}
\end{equation}
Substituting these constraints into Eq.~\eqref{2nd order SPs of NGR}, we obtain the second-order perturbed Lagrangian density as follows:
\begin{equation}
\begin{split}
    & \mathcal{L}^\mathrm{(SP\,;\,2nd\,order)}_\mathrm{total} 
    = 
    \mathcal{L}^\mathrm{(SP)}_\mathrm{NGR} + \mathcal{L}^\mathrm{(SP)}_\mathrm{matter} 
    \\
    &= ( 2 c_{1} - c_{2} + c_{3} ) a^{3} \delta^{ij} \partial_{i}{\dot{\alpha}} \partial_{j}{\dot{\alpha}}
    + 4 ( 2 c_{1} - c_{2} + 3 c_{3} ) a^{3} H \delta^{ij} \partial_{i}{\dot{\alpha}} \partial_{j}{\alpha}
    - 3 ( c_{2} - 3 c_{3} ) a^{3} H^{2} \delta^{ij} \partial_{i}{\alpha} \partial_{j}{\alpha}
    \\
    &\qquad
    + \frac{1}{2} a^{3} \dot{\Phi}_{0} \dot{\Phi}_{0} \delta^{ij} \partial_{i}{\alpha} \partial_{j}{\alpha}
    + 2 a^{2} \dot{\Phi}_{0} \delta^{ij} \partial_{i}{\alpha} \partial_{j}{\delta\Phi}
    \\
    &\qquad
    + \psi \Bigg{[}
    ( 2 c_{1} - c_{2} + c_{3} ) a \delta^{ij} \partial_{i}\partial_{j}{\psi}
    + 3 ( 2 c_{1} - c_{2} + 3 c_{3} ) a^{3} H^{2} \psi
    - \frac{1}{2} a^{3} \dot{\Phi}_{0} \dot{\Phi}_{0} \psi
    \Bigg{]}
    \\
    &\qquad
    + F \Bigg{[}
    - 10 c_{1} a^{3} H^{2} \delta^{ij} \partial_{i}\partial_{j}{F}
    + ( 2 c_{1} - c_{2} + 2 c_{3} ) a \delta^{ij} \delta^{kl} \partial_{i}\partial_{j}\partial_{k}\partial_{l}{F}
    \Bigg{]}
    \\
    &\qquad
    + \frac{1}{2} a^{3} \dot{\delta\Phi} \dot{\delta\Phi}
    - \frac{1}{2} a \delta^{ij} \partial_{i}{\delta\Phi} \partial_{j}{\delta\Phi}
    - \frac{1}{2} a^{3} V''_{0} \delta\Phi \delta\Phi
    \,.
\end{split}
\label{2nd order SPs of NGR with constraints}
\end{equation}
We find that all perturbation fields, $\alpha$, $\psi$, and $F$, decouple from each other, and the scalar mode $\alpha$ can propagate.
The ghost-free condition for the scalar mode $\alpha$ is given as
\begin{equation}
    2 c_{1} - c_{2} + c_{3} > 0 \,.
\label{ghost-free condition of lorentz scalar}
\end{equation}

For regular systems, the scalar mode $\alpha$ does not propagate in Types 2, 6 (TEGR), and 9, whereas it propagates in all other types. 
For irregular systems, this mode propagates in Types 4 and 7. 
We remark that, as mentioned in Sec.~\ref{02:01}, in irregular systems, the DOFs based on the DB analysis cannot provide the upper bound of the perturbation. 
Thus, in Type 7, the propagating mode $\alpha$ can exist.

\subsection{\label{05:03}Pseudo-scalar Perturbation}

We calculate the pseudo-scalar perturbation of NGR up to second order. 
Focusing on the pseudo-scalar terms of Eq.~\eqref{gauge fixed vierbein perturbations}, the co-vierbein, the vierbein, the metric, the inverse metric components, and the determinant of the co-vierbein are derived as follows:
\begin{equation}
\begin{split}
    e^{0}{}_{\mu}\,dx^{\mu} 
    &= 
    dt 
    \,,\\
    e^{a}{}_{\mu}\,dx^{\mu} 
    &=
    a\,\delta^{aj}\,
    \left(\delta_{ij} - \epsilon_{ijk}\,\delta^{kl}\,\partial_{l}\tilde{\sigma}\right)\,dx^{i} 
    \,,\\
    e_{0}{}^{\mu}\,\frac{\partial}{\partial x^{\mu}} 
    &=
    \frac{\partial}{\partial t} 
    \,,\\
    e_{a}{}^{\mu}\,\frac{\partial}{\partial x^{\mu}} 
    &= 
    a^{-1}\,\left[\delta_{a}{}^{i} - \delta_{ak}\,\epsilon^{ijk}\,\partial_{j}\tilde{\sigma} + \delta_{an}\,\delta_{lm}\,\epsilon^{lik}\,\epsilon^{mjn}\,\partial_{j}\tilde{\sigma}\,\partial_{k}\tilde{\sigma}\right]\,\frac{\partial}{\partial x^{i}} 
    \,,\\
    g_{\mu\nu}\,dx^{\mu}dx^{\nu} 
    &=
    - dtdt +a^{2}\,\left[\delta_{ij} + \epsilon_{mik}\,\epsilon_{ljn}\,\delta^{ml}\,\delta^{kp}\,\delta^{nq}\,\partial_{p}\tilde{\sigma}\,\partial_{q}\tilde{\sigma}\right]\,dx^{i}dx^{j} 
    \,,\\
    g^{\mu\nu}\,\frac{\partial}{\partial\,x^{\mu}}\frac{\partial}{\partial\,x^{\nu}} 
    &= 
    - \frac{\partial}{\partial t}\frac{\partial}{\partial t} + a^{-2}\,\left[\delta^{ij} - \delta_{ml}\,\epsilon^{mik}\,\epsilon^{ljn}\,\partial_{k}\tilde{\sigma}\,\partial_{n}\tilde{\sigma}\right]\,\frac{\partial}{\partial x^{i}}\frac{\partial}{\partial x^{j}} 
    \,, \\
    \theta 
    &= 
    a^{3} \left( 1 + \delta^{ij} \partial_{i}\tilde{\sigma} \partial_{j}\tilde{\sigma} \right) \,,
\end{split}
\label{vierbein and metric pseudo-SPs}
\end{equation}
where $\epsilon^{ijk}$ is the Levi-Civita symbol.\footnote{
Expanding the third term of the right-hand side of the fourth equation in Eq.~\eqref{vierbein and metric pseudo-SPs}, we get
\begin{equation}
    e_{a}{}^{\mu}\,\frac{\partial}{\partial x^{\mu}} = a^{-1}\,\left[\delta_{a}{}^{n} - \delta_{ai}\,\epsilon^{ijn}\,\partial_{j}\tilde{\sigma} +  \delta_{ai}\,\delta^{ij}\,\partial_{j}\tilde{\sigma}\,\partial_{n}\tilde{\sigma} - \delta_{an}\,\delta^{ij}\,\partial_{i}\tilde{\sigma}\,\partial_{j}\tilde{\sigma}\right]\,\frac{\partial}{\partial x^{n}}\,.
\label{}
\end{equation}
This result coincides exactly with the pioneering work Ref.~\cite{Izumi:2012qj}. However, we do not expand the Levi-Civita symbol at this stage for convenience in calculation. 
}
The torsion tensor components given in Appendix~\ref{app:03}. 

Repeating the same procedure as that of the tensor and scalar perturbation, we obtain 
\begin{equation}
\begin{split}
    &\mathcal{L}^\mathrm{(pseud\,SP\,;\,2nd\,order)}_\mathrm{total} 
    = 
    \mathcal{L}^\mathrm{(pseud\,SP)}_\mathrm{NGR} + \mathcal{L}^\mathrm{(pseud\,SP)}_\mathrm{matter} 
    \\
    &=
    - 2 ( 2 c_{1} + c_{2} ) \delta^{i j} \partial_{i}{\dot{\tilde{\sigma}}} \partial_{j}{\dot{\tilde{\sigma}}} 
    - 4 ( 2 c_{1} - c_{2} + 3 c_{3} ) H a^{3} \delta^{i j} \partial_{i}{\tilde{\sigma}} \partial_{i}{\dot{\tilde{\sigma}}} 
    \\
    &\qquad 
    + ( 2 c_{1} - c_{2} ) a \delta^{i j} \delta^{k l} \partial_{i}\partial_{k}{\tilde{\sigma}} \partial_{j}\partial_{l}{\tilde{\sigma}} 
    - 3 ( 2 c_{1} - c_{2} + 3 c_{3} ) H^{2} a^{3} \delta^{i j} \partial_{i}{\tilde{\sigma}} \partial_{j}{\tilde{\sigma}} 
    \\
    &\qquad 
    + 2 ( 2 c_{1} + 3 c_{2} ) a \delta^{i j} \delta^{k l} \partial_{i}\partial_{j}{\tilde{\sigma}} \partial_{k}\partial_{l}{\tilde{\sigma}} 
    - a^{3} V_{0} \delta^{i j} \partial_{i}{\tilde{\sigma}} \partial_{j}{\tilde{\sigma}} 
    \\
    &\qquad 
    + \frac{1}{2} a^{3} \dot{\Phi}_{0} \dot{\Phi}_{0} \delta^{i j} \partial_{i}{\tilde{\sigma}} \partial_{j}{\tilde{\sigma}} 
    + \frac{1}{2} a^{3} \dot{\delta\Phi} \dot{\delta\Phi} 
    - \frac{1}{2} a \delta^{i j} \partial_{i}{\delta\Phi} \partial_{j}{\delta\Phi} 
    - \frac{1}{2} a^{3} V''_{0} \delta\Phi \delta\Phi \,.
\end{split}
\label{pseudo-SPs of NGR with time-developing matter}
\end{equation}
The ghost-free condition for the pseudo-scalar mode $\tilde{\sigma}$ is 
\begin{equation}
    2 c_{1} + c_{2} < 0 \,,
\label{ghost-free condition of lorentz pseudo-scalar}
\end{equation}
and $c_{3}$ is arbitrary.

We find that the perturbation field $\tilde{\sigma}$ can propagate in NGR.
For regular systems, the pseudo-scalar mode does not propagate in Types 3, 6 (TEGR), and 8, whereas this mode propagates in Types 1, 2, 5, and 9. 
For irregular systems, Type 7 does not contain the pseudo-scalar mode, while Type 4 does.

\subsection{\label{05:04}Vector and pseudo-vector Perturbation}

We calculate the vector and pseudo-vector perturbations of NGR up to the second order.
We note that a coupling term between $\alpha_{i}$ and $\tilde{V}_{i}$ appears, indicating that we cannot separate these perturbations. 
Focusing on the vector and pseudo-vector terms of Eq.~\eqref{gauge fixed vierbein perturbations}, we obtain the co-vierbein, the vierbein, the metric, and the inverse metric components as follows:
\begin{equation}
\begin{split}
    e^{0}{}_{\mu}\,dx^{\mu} 
    &= 
    dt + a\,\alpha_{i}\,dx^{i} 
    \,,\\
    e^{a}{}_{\mu}\,dx^{\mu} 
    &=
    \delta^{ai}\,G_{i}\,dt + a\,\delta^{aj}\,\left(\delta_{ij} - \epsilon_{ijk}\,\delta^{kl}\,\tilde{V}_{l}\right)\,dx^{i} 
    \,,\\
    e_{0}{}^{\mu}\,\frac{\partial}{\partial x^{\mu}} 
    &= 
    \left[1 + \delta^{ij}\alpha_{i}G_{j}\right]\frac{\partial}{\partial t} 
    + a^{-1}\left[-\delta^{ij}G_{j} - \epsilon^{ijk}G_{j}\tilde{V}_{k}\right]\frac{\partial}{\partial x^{i}} 
    \,,\\
    e_{a}{}^{\mu}\,\frac{\partial}{\partial x^{\mu}} 
    &= 
    \left[-\delta_{a}{}^{i}\alpha_{i} + \delta_{aj}\epsilon^{ijk}\alpha_{i}\tilde{V}_{k}\right]\frac{\partial}{\partial t} 
    \\
    &\qquad
    + a^{-1}\left[\delta_{a}{}^{i} + \epsilon^{ijk}\delta_{aj}\tilde{V}_{k} + \delta^{ij}\delta_{a}{}^{k}G_{j}\alpha_{k} - \delta_{kl}\delta_{an}\epsilon^{kij}\epsilon^{lnm}\tilde{V}_{j}\tilde{V}_{m}\right]\frac{\partial}{\partial x^{i}}
    \,,\\
    g_{\mu\nu}\,dx^{\mu}dx^{\nu} 
    &= 
    \left[-1 + \delta^{ij}G_{i}G_{j}\right]dtdt 
    + a\left[
        -2\alpha_{i} + 2G_{i} -2\delta^{jl}\delta^{km}\epsilon_{ijk}G_{l}\tilde{V}_{m}
    \right]dtdx^{i} 
    \\
    &\qquad
    + a^{2}\left[
    \delta_{ij} - \alpha_{i}\alpha_{j} 
    + \delta^{pn}\delta^{kl}\delta^{qr}\epsilon_{pik}\epsilon_{njq}\tilde{V}_{l}\tilde{V}_{r}
    \right]dx^{i}dx^{j}
    \,, \\
    g^{\mu\nu}\,\frac{\partial}{\partial\,x^{\mu}}\frac{\partial}{\partial\,x^{\nu}} 
    &= 
    \left[-1 -2\delta^{ij}\alpha_{i}G_{j} + \delta^{ij}\alpha_{i}\alpha_{j}\right]
    \frac{\partial}{\partial t}\frac{\partial}{\partial t} 
    \\
    &\qquad
    + a^{-1}\left[
        2\delta^{ij}G_{j} - 2\delta^{ij}\alpha_{j} + 2\epsilon^{ijk}G_{j}\tilde{V}_{k}
    \right]
    \frac{\partial}{\partial t}\frac{\partial} {\partial x^{i}} 
    \\
    &\qquad
    + a^{-2}\left[
        \delta^{ij} - \delta^{ik}\delta^{jl}G_{k}G_{l} 
        + 2 \delta^{ik}\delta^{jl}\alpha_{k}G_{l} 
        - \delta_{lm}\epsilon^{lik}\epsilon^{mjn}\tilde{V}_{k}\tilde{V}_{n}
    \right]
    \frac{\partial} {\partial x^{i}}\frac{\partial} {\partial x^{j}} 
    \,, \\
    \theta 
    &= 
    a^{3} \left( 1 + \delta^{ij}\,\tilde{V}_{i}\,\tilde{V}_{j} -\delta^{ij}\alpha_{i}\,G_{j} \right)
    \,,
\end{split}
\label{vierbein and metric in VPs}
\end{equation}
The torsion tensor components given in Appendix~\ref{app:04}. 

Repeating the same procedure, we obtain 
\begin{equation}
\begin{split}
    &\mathcal{L}^\mathrm{(VP\,and\,pseudo\,VP\,;\,2nd\,order)}_\mathrm{total} 
    = 
    \mathcal{L}^\mathrm{(VP\,and\,pseudo\,VP)}_\mathrm{NGR} + \mathcal{L}^\mathrm{(VP\,and\,pseudo\,VP)}_\mathrm{matter} 
    \\
    &= 
    ( 2 c_{1} - c_{2} + c_{3} ) a^{3} \delta^{ij} \dot{\alpha}_{i} \dot{\alpha}_{j} 
    + 4 ( 2 c_{1} - c_{2} + 3 c_{3} ) a^{3} H \delta^{ij} \dot{\alpha}_{i} \alpha_{j} \\
    &\qquad 
    - 2 c_{1} a \delta^{ij} \delta^{kl} \partial_{i}{\alpha_{k}} \partial_{j}{\alpha_{l}} 
    - 3 ( c_{2} - 3 c_{3} ) a^{3} H^{2} \delta^{ij} \alpha_{i} \alpha_{j} 
    + 2 a^{2} \dot{\Phi}_{0} \delta^{ij} \alpha_{i} \partial_{j}{\delta\Phi} \\
    &\qquad 
    - 2 ( 2 c_{1} + c_{2} ) a^{3} \delta^{ij} \dot{\tilde{V}}_{i} \dot{\tilde{V}}_{j} 
    - 4 ( 2 c_{1} - c_{2} + 3 c_{3} ) a^{3} H \delta^{ij} \dot{\tilde{V}}_{i} \tilde{V}_{j} 
    - a^{3} V_{0} \delta^{ij} \tilde{V}_{i} \tilde{V}_{j} 
    \\
    &\qquad 
    + ( 2 c_{1} - c_{2} + c_{3} ) a \delta^{ij} \delta^{kl} \partial_{i}{\tilde{V}_{k}} \partial_{j}{\tilde{V}_{l}} 
    - 3 ( 2 c_{1} - c_{2} + 3 c_{3} ) a^{3} H^{2} \delta^{ij} \tilde{V}_{i} \tilde{V}_{j} 
    \\
    &\qquad 
    + 2 ( 2 c_{2} - c_{3} ) a^{2} \epsilon^{ijk} \partial_{i}{\dot{\alpha}_{j}} \tilde{V}_{k} 
    - 4 ( 2 c_{1} - c_{2} + 3 c_{3} ) a^{2} H \epsilon^{ijk} \alpha_{i} \partial_{j}{\tilde{V}_{k}} 
    \\
    &\qquad 
    + G_{i} \Bigg{[}
    - 2 c_{1} a \delta^{ij} \delta^{kl} \partial_{k}\partial_{l}{G_{j}} 
    + 6 c_{1} a^{3} H^{2} \delta^{ij} G_{j} \Bigg{]} 
    \\
    &\qquad
    + \frac{1}{2} a^{3} \dot{\delta\Phi} \dot{\delta\Phi} 
    - \frac{1}{2} a \delta^{ij} \partial_{i}{\delta\Phi} \partial_{j}{\delta\Phi}
    - \frac{1}{2} V''_{0} a^{3} \delta\Phi \delta\Phi \,.
\end{split}
\label{simplified VPs and pseudo-VPs of NGR with time-developing matter}
\end{equation}
We notice that the perturbation field $G_{i}$ decouples from other fields, but $\alpha_{i}$ and $\tilde{V}_{i}\,$ are coupled with each other as in the sixth line of Eq.~\eqref{simplified VPs and pseudo-VPs of NGR with time-developing matter}. 
The ghost-free conditions for the vector and pseudo-vector modes, $\alpha_{i}$ and $\tilde{V}_{i}$, are given as
\begin{equation}
    2 c_{1} - c_{2} + c_{3} > 0 \,,
\label{ghost-free condition of lorentz vector}
\end{equation}
and 
\begin{equation}
    2 c_{1} + c_{2} < 0\qquad \mathrm{and} \qquad \mbox{$c_3$ is arbitrary} \,
\label{ghost-free condtion of lorentz pseudo-vecto}
\end{equation}
respectively.

First, we consider regular systems. (See Table~\ref{Table:fullDoFofNGR}.)
In the case $2 c_{1} - c_{2} + c_{3}=0$ and $2 c_{1} + c_{2}=0$, that is, in Type 6 (TEGR), we find $\alpha$ and $\tilde{V}_{i}$ do not propagate, as desired. 

In the case solely $2 c_{1} - c_{2} + c_{3} = 0\,$, that is, in Types 2 and 9, the perturbed Lagrangian density turns into as follows:
\begin{equation}
\begin{split}
    &\mathcal{L}^\mathrm{(VP\,and\,pseudo\,VP)}_\mathrm{total} 
    =
    \mathcal{L}^\mathrm{(VP\,and\,pseudo\,VP)}_\mathrm{NGR} + \mathcal{L}^\mathrm{(VP\,and\,pseudo\,VP)}_\mathrm{matter} 
    \\
    &= 
    - 2 ( 2 c_{1} + c_{2} ) a^{3} \delta^{ij} \dot{\tilde{V}}_{i} \dot{\tilde{V}}_{j} 
    - 4 ( 2 c_{1} - c_{2} + 3 c_{3} ) a^{3} H \delta^{ij} \dot{\tilde{V}}_{i} \tilde{V}_{j} 
    - a^{3} V_{0} \delta^{ij} \tilde{V}_{i} \tilde{V}_{j}
    \\
    &\qquad
    - 3 ( 2 c_{1} - c_{2} + 3 c_{3} ) a^{3} H^{2} \delta^{ij} \tilde{V}_{i} \tilde{V}_{j} 
    \\
    &\qquad 
    + \alpha_{i} \Bigg{[} 
    4 ( 2 c_{1} - c_{2} + 3 c_{3} ) a^{3} H \delta^{ij} \dot{\alpha}_{j} 
    + 12 ( 2 c_{1} - c_{2} + 3 c_{3} ) a^{3} H^{2} \delta^{ij} \alpha_{j} 
    + 4 ( 2 c_{1} - c_{2} + 3 c_{3} ) a^{3} \dot{H} \delta^{ij} \alpha_{j} 
    \\
    &\qquad \qquad \qquad
    - 2 c_{1} a \delta^{ij} \delta^{kl} \partial_{k}\partial_{l}{\alpha_{j}}
    + 3 ( c_{2} - 3 c_{3} ) a^{3} H^{2} \delta^{ij} \alpha_{j} 
    \Bigg{]} 
    \\
    &\qquad 
    + G_{i} \Bigg{[} 
    - 2 c_{1} a \delta^{ij} \delta^{kl} \partial_{k}\partial_{l}{G_{j}}
    + 6 c_{1} a^{3} H^{2} \delta^{ij} G_{j} 
    \Bigg{]} 
    \\
    &\qquad 
    + \frac{1}{2} a^{3} \dot{\delta\Phi} \dot{\delta\Phi} 
    - \frac{1}{2} a \delta^{ij} \partial_{i}{\delta\Phi} \partial_{j}{\delta\Phi} 
    - \frac{1}{2} V''_{0} a^{3} \delta\Phi \delta\Phi 
    \,,
\end{split}
\label{further simplified VPs and pseudo-VPs of NGR with time-developing matter in Type 2}
\end{equation}
where we used the constraint with respect to $\alpha_{i}$.  
We find that all the perturbation modes decouple from each other, and the pseudo-vector mode $\tilde{V}_{i}$ propagates in Types 2 and 9. 

In the case solely $2 c_{1} + c_{2} = 0\,$, that is, in Types 3 and 8, the perturbed Lagrangian density is 
\begin{equation}
\begin{split}
    &\mathcal{L}^\mathrm{(VP\,and\,pseudo\,VP)}_\mathrm{total} 
    =
    \mathcal{L}^\mathrm{(VP\,and\,pseudo\,VP)}_\mathrm{NGR} + \mathcal{L}^\mathrm{(VP\,and\,pseudo\,VP)}_\mathrm{matter} 
    \\
    &= 
    ( 2 c_{1} - c_{2} + c_{3} ) a^{3} \delta^{ij} \dot{\alpha}_{i} \dot{\alpha}_{j} + 4 ( 2 c_{1} - c_{2} + 3 c_{3} ) a^{3} H \delta^{ij} \dot{\alpha}_{i} \alpha_{j} 
    \\
    &\qquad 
    - 2 c_{1} a \delta^{ij} \delta^{kl} \partial_{i}{\alpha_{k}} \partial_{j}{\alpha_{l}} 
    - 3 ( c_{2} - 3 c_{3} ) a^{3} H^{2} \delta^{ij} \alpha_{i} \alpha_{j}
    + 2 a^{2} \dot{\Phi}_{0} \delta^{ij} \alpha_{i} \partial_{j}{\delta\Phi} 
    \\
    &\qquad 
    + \tilde{V}_{i} \Bigg{[}
    - 4 ( 2 c_{1} - c_{2} + 3 c_{3} ) a^{3} H \delta^{ij} \dot{\tilde{V}}_{j} 
    - 12 ( 2 c_{1} - c_{2} + 3 c_{3} ) a^{3} H^{2} \dot{\tilde{V}}_{j} 
    \\
    &\qquad \qquad \quad
    - 4 ( 2 c_{1} - c_{2} + 3 c_{3} ) a^{3} \dot{H} \delta^{ij} \tilde{V}_{j} 
    +  a^{3} V_{0} \delta^{ij} \tilde{V}_{j} 
    \\
    &\qquad \qquad \quad
    + ( 2 c_{1} - c_{2} + c_{3} ) a \delta^{ij} \delta^{kl} \partial_{k}\partial_{l}{\tilde{V}_{j}} 
    + 3 ( 2 c_{1} - c_{2} + 3 c_{3} ) a^{3} H^{2} \delta^{ij} \tilde{V}_{j} 
    \Bigg{]}
    \\
    &\qquad 
    + G_{i} \Bigg{[} 
    - 2 c_{1} a \delta^{ij} \delta^{kl} \partial_{k}\partial_{l}{G_{j}} 
    + 6 c_{1} a^{3} H^{2} \delta^{ij} G_{j} \Bigg{]} 
    \\
    &\qquad 
    + \frac{1}{2} a^{3} \dot{\delta\Phi} \dot{\delta\Phi} 
    - \frac{1}{2} a \delta^{ij} \partial_{i}{\delta\Phi} \partial_{j}{\delta\Phi} 
    - \frac{1}{2} V''_{0} a^{3} \delta\Phi \delta\Phi 
    \,,
\end{split}
\label{further simplified VPs and pseudo-VPs of NGR with time-developing matter in Type 3 and Type 7}
\end{equation}
where we used the constraint with respect to $\tilde{V}_{i}$. 
We find that all perturbation modes decouple from each other, and the vector mode $\alpha_{i}$ propagates in Types 3 and 8.

In Type 1, the parameters $c_{1}$, $c_{2}$, and $c_{3}$ can be freely chosen. 
This flexibility allows us to select values for $c_{1}$, $c_{2}$, and $c_{3}$ such that they satisfy both $2c_{2} - c_{3} = 0$ and $2c_{1} - c_{2} + 3c_{3} = 0$, which is valid for decoupling $\alpha$ and $\tilde{V}_{i}$ modes.
Then, all perturbation modes decouple from each other, the vector mode $\alpha_{i}$ and the pseudo-vector mode $\tilde{V}_{i}$ propagate. 
We conclude that Type 1 can contain at most the vector and pseudo-vector modes. 
In Type 5, however, we cannot make the perturbation modes $\alpha_{i}$ and $\tilde{V}_{i}$ decouple due to the existence of the first term in the sixth line of Eq.~\eqref{simplified VPs and pseudo-VPs of NGR with time-developing matter}. 
Taking into account the result of the DB analysis~\cite{Tomonari:2024ybs}, which suggests the upper bound of DOFs in Type 5 is seven, we can conclude that either $\alpha_{i}$ or $\tilde{V}_{i}$ propagates. 

Second, we consider irregular systems. (See Table~\ref{Table:fullDoFofNGR}.)

Type 7 contains the vector mode $\alpha_{i}$ since this type satisfies the common condition with that of Type 3. 
In Type 4, unfortunately, we cannot decouple any modes in Eq.~\eqref{simplified VPs and pseudo-VPs of NGR with time-developing matter}. 
In Ref.~\cite{Tomonari:2024lpv}, it is shown that the constraint surface, denote it by $\Gamma_{0}$, of Type 4 contains that of its regularized system, denote it by $\Gamma_{1}$, which implies that a part of the perturbation modes could be ascribed to the DOFs of the outside region of the regularized system $\Gamma_{0} \cap {}^{\lnot}\Gamma_{1}$. 
If this is the case, Type 4 contains either the vector mode $\alpha_{i}$ or pseudo-vector mode $\tilde{V}_{i}\,$.

\section{\label{06}Conclusions}

In this work, we have investigated the propagating mode in each type of NGR up to second order. 
After summarizing recent progress on the Hamiltonian analysis of NGR, we reconsider the vierbein perturbation framework and clarified the correspondence between each perturbation field and vierbein component. 
Throughout this, we addressed the issues 1) and 2) in Sec.~\ref{01}. 
We revealed that the spatially flat gauge is an adequate gauge choice in a theory with the violation of local LI, which addresses issue 3) in Sec.~\ref{01}. 
To consider cosmological perturbations, we introduced a scalar field as matter and derived the background-field equations of NGR. 
Finally, we performed the perturbative analysis of NGR up to second order to reveal the propagating modes in each type of NGR. 
The results are summarized in Table~\ref{Table:CosmoPertOfNGR}. 
The emergence of modes is consistent with the consideration in Sec.~III-F of Ref.~\cite{Tomonari:2024ybs} and Sec.~III-F of Ref.~\cite{Tomonari:2024lpv}.
We found that new propagating modes in the second-order perturbation theory of NGR, which addresses issue 4) in Sec.~\ref{01}.
\begin{table}[ht!]
    \centering
    \renewcommand{\arraystretch}{1.5}
    \begin{tabular}{ c || c | c | c | c }
	Theory & Regularity & 
    \begin{tabular}{c}
        \# of Non-linear DOF \\ 
        (DB analysis)
    \end{tabular} 
    & 
    \begin{tabular}{c}
        Propagating Modes \\ 
        (Perturbative analysis) 
    \end{tabular}  
    & Ghost-free conditions\\ \hline \hline
	Type 1 & $-$ & 8 & 6 - 8: ($h_{ij}, \alpha, \tilde{\sigma}, \alpha_{i} ($or $\tilde{V}_{i})$; $\tilde{V}_{i}($or $\alpha_{i})$) & \begin{tabular}{c} $2 c_{1} - c_{2} < 0$ \& \\ $2 c_{1} - c_{2} + c_{3} > 0$ \& \\ $2 c_{1} + c_{2} < 0$ \end{tabular} \\ \hline
	Type 2 & $\checkmark$ & 6 & 5: ($h_{ij}, \tilde{\sigma}, \tilde{V}_{i}$) & \begin{tabular}{c} $2 c_{1} - c_{2} < 0$ \& \\ $2 c_{1} + c_{2} <0$ \end{tabular} \\ \hline
	Type 3 & $\checkmark$ & 5 & 5: ($h_{ij}, \alpha, \alpha_{i}$) & \begin{tabular}{c} $2 c_{1} - c_{2} < 0$ \& \\ $2 c_{1} - c_{2} + c_{3} > 0$	\end{tabular}\\ \hline
	Type 4 & $\times$ & 5 & 2 - 4: ($\alpha, \tilde{\sigma}$; either $\alpha_{i}$ or $\tilde{V}_{i}$) & \begin{tabular}{c} $2 c_{1} - c_{2} + c_{3} > 0$ \& \\ $2 c_{1} + c_{2} < 0$ \end{tabular}\\ \hline
	Type 5 & $\checkmark$ & 7 & 6: ($h_{ij}, \alpha, \tilde{\sigma},$ either $\alpha_{i}$ or $\tilde{V}_{i}$) & \begin{tabular}{c} $2 c_{1} - c_{2} < 0$ \& \\ $2 c_{1} - c_{2} + c_{3} > 0$ \& \\ $2 c_{1} + c_{2} < 0$ \end{tabular}\\ \hline
	Type 6 & $\checkmark$ & 2 & 2: ($h_{ij}$) & $2 c_{1} - c_{2} < 0$ \\ \hline
	Type 7 & $\times$ & 0 (Topological) & 3: ($\alpha, \alpha_{i}$) & \begin{tabular}{c} $2 c_{1} - c_{2} + c_{3} > 0$ \end{tabular}\\ \hline
	Type 8 & $\checkmark$ & 6 or 4 (Bifurcate) & 5: ($h_{ij}, \alpha, \alpha_{i}$) & \begin{tabular}{c} $2 c_{1} - c_{2} < 0$ \& \\ $2 c_{1} - c_{2} + c_{3} > 0$	\end{tabular}\\ \hline
	Type 9 & $\checkmark$ & 3 & 3: ($\tilde{\sigma}, \tilde{V}_{i}$) & $2 c_{1} + c_{2} < 0$ \\ \hline
    \end{tabular}
    \caption{
    We summarize our work on the linear perturbations of NGR around the flat FLRW background spacetime in the above table.
    Perturbations that are listed on the left-hand side to `` $;$ '' always propagate; Perturbations given on the right-hand side to `` $;$ '' can propagate under the imposition of a specific condition on the parameters. 
    We denote ``XXX - YYY'' by the meaning of ranging from XXX to YYY and an invalid case, respectively. 
    Type 6 is TEGR, which is equivalent to GR.
    We remark that for irregular systems, the non-linear DOF cannot restrict the number of the perturbation modes. See Sec.~\ref{02:01} or Refs.~\cite{Miskovic:2003ex,Tomonari:2024lpv} in detail.  
    }
    \label{Table:CosmoPertOfNGR}
\end{table}

We compare our result with the previous works~\cite{Golovnev:2023ddv, Golovnev:2023jnc, Bahamonde:2024zkb}. 
The perturbative analysis around the Minkowski background has been performed by several groups in Refs.~\cite{Golovnev:2023ddv, Bahamonde:2024zkb}. 
In Ref.~\cite{Bahamonde:2024zkb}, Types 1, 2, 3, 5, 6 (TEGR), and 8 are considered gravitational theories including tensorial propagating modes.
The results of this work coincide with those of Ref.~\cite{Golovnev:2023ddv} except for Type 8; it concluded that the tensor mode does not exist.
In Ref.~\cite{Golovnev:2023jnc}, applying the conformal transformation to the result in Ref.~\cite{Golovnev:2023ddv}, cosmological perturbative analysis is carried out. 
In all cases, no new propagating modes appear; pure gauge degrees of freedom are converted into constrained variables.
Our result differs from Ref.~\cite{Golovnev:2023jnc}, except for Types 1 and 6 (TEGR). 
We shall enumerate the differences as follows:\footnote{
For the reader's convenience, we list the correspondence of perturbation fields between our work and Ref.~\cite{Golovnev:2023ddv} as follows (the left variables are ours, the right variables are those in Ref.~\cite{Golovnev:2023ddv}):
\begin{equation}
\begin{split}
    &
    \varphi \leftrightarrow \psi\,, 
    \quad 
    \psi \leftrightarrow \phi\,, 
    \quad 
    B \leftrightarrow \sigma\,, 
    \quad 
    F \leftrightarrow \zeta\,, 
    \quad 
    \alpha \leftrightarrow \beta \,, 
    \quad 
    \tilde{\sigma} \leftrightarrow s \,, \\
    &
    C_{i} \leftrightarrow c_{i} \,,
    \quad
    G_{i} \leftrightarrow v_{i} \,, 
    \quad 
    \alpha_{i} \leftrightarrow u_{i} \,, 
    \quad 
    \tilde{V}_{i} \leftrightarrow \chi_{i} \,.
\end{split}
\end{equation}
The tensor mode is represented using the standard notation. 
In our formulation, $C_{i}$ is introduced as $2 \partial_{(i}{C_{j)}}$ according to the convention, by contrast, they introduce it just as $c_{i}$. 
Moreover, in our analysis, the perturbation field $G_{i}$ does not propagate, which is a constrained variable.
Thus, we can regard the perturbation field $\alpha_{i}$ in our notation as both $\mathcal{M}_{i} = ( u_{i} - v_{i} ) / 2$ and $\mathcal{L}_{i} = ( u_{i} + v_{i} ) / 2\,$ in their notation. 
}
\begin{itemize}
    \item Type 1  \\
    There is almost no difference from~\cite{Golovnev:2023jnc} except for the vector and pseudo-vector modes, $\alpha_{i}$ and $\tilde{V}_{i}$, that decouple only in a specific case.
    Ghost-free parameter space exists, and our result differs from that in Ref.~\cite{Bahamonde:2024zkb}.
    \item Type 2 \\
    The pseudo-vector mode $\tilde{V}_{i}$ propagates in our case, in addition to the propagating modes in Ref.~\cite{Golovnev:2023jnc}. 
    Ghost-free parameter space exists, and our result coincides with that in Ref.~\cite{Bahamonde:2024zkb}.
    \item Type 3 \\
    The vector mode $\alpha_{i}$ propagates in our case, in addition to the propagating modes in Ref.~\cite{Golovnev:2023jnc}. 
    Ghost-free parameter space exists, and our result coincides with that in Ref.~\cite{Bahamonde:2024zkb}.
    \item Type 4 \\
    In our case, the scalar mode $\alpha\,$ propagates.
    Either the vector or pseudo-vector modes, $\alpha_{i}$ or $\tilde{V}_{i}$, can propagate in a specific case, whereas only half of $\alpha_{i}\,$ always propagates in Ref.~\cite{Golovnev:2023jnc}. 
    The pseudo-scalar mode $\tilde{\sigma}$ propagates as in Ref.~\cite{Golovnev:2023jnc}.
    Ghost-free parameter space exists.
    \item Type 5 \\
    The scalar mode $\alpha$ propagates in our case.
    However, the vector mode $\alpha_{i}$ propagates in a specific case, and if this mode propagates, the pseudo-vector mode $\tilde{V}_{i}$ cannot propagate, and vice versa.
    Both the scalar and vector modes propagate in Ref.~\cite{Golovnev:2023jnc}.
    The propagation of the tensor and pseudo-scalar modes, $h_{ij}$ and $\tilde{\sigma}\,$, is the same as in Ref.~\cite{Golovnev:2023jnc}.
    Ghost-free parameter space exists, and our result differs from that in Ref.~\cite{Bahamonde:2024zkb}.
    \item Type 6 (TEGR) \\
    There is no difference from Ref.~\cite{Golovnev:2023jnc}.
    Ghost-free parameter space exists, and our result coincides with that in Ref.~\cite{Bahamonde:2024zkb}.
    \item Type 7 \\
    The scalar and vector modes, $\alpha$ and $\alpha_{i}$, propagate, but the tensor mode $h_{ij}$ does not in our analysis. 
    On the other hand, only the tensor mode $h_{ij}$ propagates in Ref.~\cite{Golovnev:2023jnc}. 
    Ghost-free parameter space exists.
    \item Type 8 \\
    The scalar, vector, and tensor modes, $\alpha$, $\alpha_{i}$, and $h_{ij}$, propagate in our case, whereas no propagating mode exists in Ref.~\cite{Golovnev:2023jnc}.
    Ghost-free parameter space exists, and our result coincides with that in Ref.~\cite{Bahamonde:2024zkb}.
    \item Type 9 \\
    The pseudo-vector mode $\tilde{V}_{i}$ propagates in our case, in addition to the propagating modes in Ref.~\cite{Golovnev:2023jnc}. 
    Ghost-free parameter space exists.
\end{itemize}

The additional propagation modes in Types 2, 3, and 9 can be attributed to higher-order perturbative terms included in our analysis. 
By contrast, the differences in the propagating modes of Types 4, 5, 7, and 8 between Ref.~\cite{Golovnev:2023jnc} and our work may stem from different gauge choices. 
It should be noted that these analyses are not carried out in terms of gauge-invariant variables.
Here, we note again that a violation of symmetry limits the proper choice of gauge.
A perturbation field originating from a broken symmetry should not be fixed, and such variables should not be confused with other perturbation fields that respect a symmetry.
For instance, one should not impose a gauge such as $\tilde{V}_{i} = 0$ or $\alpha = F$, since $F$ obeys diffeomorphism invariance, which holds in NGR, whereas $\tilde{V}_{i}$ and $\alpha$ are related to local Lorentz invariance, which is not preserved in NGR. 
Otherwise, such discrepancies may be signal issues in the perturbative framework. 
A more detailed investigation of this point is left for future work. 

In Table~\ref{Table:CosmoPertOfNGR}, we observe discrepancies between the columns \# of Non-linear DOF and Propagating mode.
There are two possible scenarios that explain these discrepancies.
The first possibility is that the linearization process accidentally restores part of the symmetry.
In this case, additional first-class constraints appear, and as a result, the total number of DOFs is reduced.
The second possibility is that strong coupling occurs.
In this case, the kinetic term responsible for the propagation of a mode may arise only at higher order in perturbation theory, thereby resolving the apparent absence of the propagating mode at the linear level.
Clarifying which of these possibilities explains the discrepancies in each type is an important issue for future work. 

Most importantly, in cosmological applications, it has been shown that Type 3 still preserves $SO(3)$ invariance. (for instance, see Refs.~\cite{Tomonari:2024lpv, Tomonari:2024ybs}.)
The preservation of $SO(3)$ invariance in Type 3 implies that the modes corresponding to this symmetry never propagate at the nonlinear level.
Our current results based on the perturbative approach are consistent with this picture and also with existing works~\cite{Golovnev:2023ddv, Golovnev:2023jnc, Bahamonde:2024zkb}, independent of any concerns regarding an improper gauge choice or potential issues in the perturbative framework. 
Furthermore, in Ref.~\cite{Bahamonde:2024zkb}, Type 3 allows parameter ranges for $c_{1}$, $c_{2}$, and $c_{3}$ that render the theory stable in the Minkowski background spacetime.
The same property can be expected in cosmological perturbations, since, according to Refs.~\cite{Golovnev:2023ddv, Golovnev:2023jnc}, a proper conformal transformation connects the results of perturbative analysis around the Minkowski background to those of the flat FLRW background spacetime. 
In our analysis, Type 3 has a ghost-free region in the parameter space; thus, our perspectives align with each other at all points.

Given that the number of DOFs in the DB and perturbative analyses coincide and that Type 3 contains the propagating tensor modes, if Type 3 does not suffer from strong couplings, a healthy MAG theory can be obtained for cosmological applications.
If this is the case, the theory will provide a new perspective on the large-scale structure formation, including dark matter issues due to the violation of boost invariance, whereas it retains the properties of isotropy in the cosmic microwave background by virtue of the $SO(3)$ invariance of the theory.
However, if this is not the case, we should investigate the possibility of implementing the screening mechanism~\cite{Brax:2013ida, Brax:2021wcv} to remedy the strong couplings, with the aim of applying it to astrophysics. 
For example, violations of the local LI could leave observable imprints in gravitational wave signals~\cite{Ghosh:2023xes}.
These issues are required for further investigation in future work.

\acknowledgments
K.T. thanks the Interfaculty Initiative in Information Studies, Graduate School of Interdisciplinary Information Studies, The University of Tokyo, for supporting this work.
K.T. thanks Cadabra for completing all calculations of this work. 
T.K. is supported by the National Science Foundation of China (No.~12403003), and the National Key R\&D Program of China (No.~2021YFA0718500).

\appendix
\section{\label{app:01}Torsion tensor in tensor perturbation}

We calculate the torsion tensor components up to second order as follows:
\begin{align}
\label{TPs}
\begin{split}
    &T^{0}{}_{0i} = 0 
    \,,\\
    &T^{0}{}_{ij} = 0 
    \,,\\
    &T^{i}{}_{0j} = H \delta^{i}{}_{j} + \delta^{ik} \dot{h}_{kj} - h^{ik} \dot{h}_{kj} 
    \,,\\
    &T^{i}{}_{jk} = - \delta^{il} \partial_{k}h_{lj} + \delta^{il} \partial_{j}h_{lk} + h^{il} \partial_{k}h_{lj} - h^{il}\,\partial_{j}h_{lk}
    \,,\\
    &T^{\mu}{}_{0\mu} = 3 H + \delta^{ij} \dot{h}_{ij} - h^{ij} \dot{h}_{ij} 
    \,,\\
    &T^{\mu}{}_{i\mu} = - \delta^{jk} \partial_{j}{h_{ik}} + \delta^{jk} \partial_{i}{h_{jk}} + h^{jk} \partial_{j}{h_{ik}} - h^{jk} \partial_{i}{h_{jk}} 
    \,,
\end{split}
\end{align}
where $h$ is the trace of $h_{ij}\,$. We set $h^{i}{}_{j} = h_{j}{}^{i} = \delta^{ik} h_{kj}$ and $h^{i}{}_{j} = h_{j}{}^{i} = \delta_{jk} h^{ik}\,$.

\section{\label{app:02}Torsion tensor in scalar perturbation}

We calculate the torsion tensor components up to second order as follows:
\begin{align}
\label{SPs}
\begin{split}
    &T^{0}{}_{0i} = 
    a \left( \partial_{i}{\dot{\alpha}} 
    - \psi \partial_{i}{\dot{\alpha}} \right) 
    - \partial_{i}{\psi} 
    + \delta^{jk} \partial_{j}{\alpha} \partial_{k}\partial_{i}{F} 
    + \psi \partial_{i}{\psi} 
    \,,\\
    &T^{0}{}_{ij} = 0 
    \,,\\
    &T^{i}{}_{0j} = 
    H \delta^{i}{}_{j} 
    - a^{-1} \delta^{ij} \left( \partial_{i}\partial_{j}{F} 
    + \partial_{i}{F} \partial_{j}{\psi} \right)
    + 2 H \delta^{ij} \partial_{i}{F} \partial_{j}{\alpha}
    + \delta^{ij} \partial_{i}{F} \partial_{j}{\dot{\alpha}}
    \,,\\
    &T^{i}{}_{jk} = 0 
    \,,\\
    &T^{\mu}{}_{0\mu} = 
    3 H
    - a^{-1} \delta^{ij} \partial_{i}\partial_{j}{F} 
    - a^{-1} \delta^{ij} \partial_{i}{F} \partial_{j}{\psi} 
    + 2 H \delta^{ij} \partial_{i}{F} \partial_{j}{\alpha}  
    + \delta^{ij} \partial_{i}{F} \partial_{j}{\dot{\alpha}}
    \,,\\
    &T^{\mu}{}_{i\mu} = 
    \partial_{i}{\psi} 
    - \delta^{ij} \partial_{i}{\alpha} \partial_{j}{F}
    - \psi \partial_{i}{\psi}
    + a \left( - \partial_{i}{\dot{\alpha}} 
    + \psi \partial_{i}{\dot{\alpha}} \right)
    \,.
\end{split}
\end{align}

\section{\label{app:03}Torsion tensor in pseudo-scalar perturbation}

We calculate the torsion tensor components up to second order as follows:
\begin{equation}
\begin{split}
    &T^{0}{}_{0i} = 0 \,,\\
    &T^{0}{}_{ij} = 0 \,,\\
    &T^{i}{}_{0j} = 
    H \delta^{i}_{j} 
    - \delta^{ik} \delta^{lm} \epsilon_{jkl} \partial_{m}{\dot{\tilde{\sigma}}} 
    - \delta^{ik} \partial_{j}{\tilde{\sigma}} \partial_{k}{\dot{\tilde{\sigma}}} 
    + \delta^{i}{}_{j} \delta^{kl} \partial_{k}{\tilde{\sigma}} \partial_{l}{\dot{\tilde{\sigma}}} 
    \, , \\
    &T^{i}{}_{jk} = 
    \delta^{il} \delta^{mn} \epsilon_{jlm} \partial_{n}\partial_{k}{\tilde{\sigma}} 
    - \delta^{il} \delta^{mn} \epsilon_{klm} \partial_{n}\partial_{j}{\tilde{\sigma}} 
    + \delta^{il} \partial_{j}{\tilde{\sigma}} \partial_{l}\partial_{k}{\tilde{\sigma}} 
    \\
    & \qquad \qquad
    - \delta^{il} \partial_{k}{\tilde{\sigma}} \partial_{l}\partial_{j}{\tilde{\sigma}} 
    - \delta^{i}{}_{j} \delta^{lm} \partial_{l}{\tilde{\sigma}} \partial_{m}\partial_{k}{\tilde{\sigma}} 
    + \delta^{i}{}_{k} \delta^{lm} \partial_{l}{\tilde{\sigma}} \partial_{m}\partial_{j}{\tilde{\sigma}} \,,\\
    &T^{\mu}{}_{0\mu} = 
    3 H + 2 \delta^{ij} \partial_{i}{\tilde{\sigma}} \partial_{j}{\dot{\tilde{\sigma}}} \,,\\
    &T^{\mu}{}_{i\mu} = 
    \delta^{jl} \delta^{km} \epsilon_{ijk} \partial_{l}\partial_{m}{\tilde{\sigma}} 
    + \delta^{kl} \partial_{i}{\tilde{\sigma}} \partial_{k}\partial_{l}{\tilde{\sigma}} 
    + \delta^{jk} \partial_{j}{\tilde{\sigma}} \partial_{k}\partial_{i}{\tilde{\sigma}} \,,
\end{split}
\label{pseudoSPs}
\end{equation}
where $\epsilon_{ijk}$ is the Levi-Civita symbol.

\section{\label{app:04}Torsion tensor in vector and pseudo-vector perturbation}

We calculate the torsion tensor components up to second order as follows:
\begin{equation}
\begin{split}
    &T^{0}{}_{0i} = 
    a \dot{\alpha}_{i} 
    + \delta^{jk} \alpha_{j} \partial_{i}{G}_{k} 
    + a \delta^{jk} \delta^{lm} \epsilon_{ijl} \alpha_{k} {\dot{\tilde{V}}_{m}} 
    \,,\\
    &T^{0}{}_{ij} = 
    a \left( \partial_{i}{\alpha_{j}} - \partial_{j}{\alpha_{i}} \right) 
    + a \left( \delta^{kl} \delta^{mn} \epsilon_{jkm} \alpha_{l} \partial_{i}{\tilde{V}_{n}} 
    - \delta^{kl} \delta^{mn} \epsilon_{ikm} \alpha_{l} \partial_{j}{\tilde{V}_{n}} \right) 
    \,,\\
    &T^{i}{}_{0j} = 
    H \delta^{i}{}_{j} 
    - a^{-1} \delta^{ik} \partial_{j}{G}_{k}
    + \delta^{il} \delta^{km} \epsilon_{ljk} \dot{\tilde{V}}_{m}
    + a^{-1} \delta^{ik} \delta^{lm} \delta^{pq} \epsilon_{klp} \tilde{V}_{q} \partial_{j}{G_{m}} 
    \\
    & \qquad \qquad
    - \delta^{ik} G_{k} \dot{\alpha}_{j} 
    + \delta^{kl} \delta^{im} \delta^{pq} \delta^{rs} \epsilon_{jkp} \epsilon_{mlr} \dot{\tilde{V}}_{q} \tilde{V}_{s}
    \,,\\
    &T^{i}{}_{jk} = 
    \delta^{li} \delta^{mn} \epsilon_{jlm} \partial_{k}{\tilde{V}_{n}} 
    - \delta^{li} \delta^{mn} \epsilon_{klm} \partial_{j}{\tilde{V}_{n}}
    + \delta^{il} G_{l} \partial_{k}{\alpha_{j}} 
    - \delta^{il} G_{l} \partial_{j}{\alpha_{k}}  
    \\
    &\qquad \qquad \quad
    - \delta^{il} \delta^{mn} \delta^{pq} \delta^{rs} \epsilon_{lnp} \epsilon_{jmr} \tilde{V}_{q} \partial_{k}{\tilde{V}_{s}}
    + \delta^{il} \delta^{mn} \delta^{pq} \delta^{rs} \epsilon_{lnp} \epsilon_{kmr} \tilde{V}_{q} \partial_{j}{\tilde{V}_{s}}
    \,,\\
    &T^{\mu}{}_{0\mu} = 
    3 H 
    - a^{-1} \delta^{ij} \partial_{i}{G_{j}} 
    - a^{-1} \delta^{ij} \delta^{kl} \delta^{mn} \epsilon_{ikm} \partial_{j}{G_{n}} \tilde{V}_{l} 
    \\
    & \qquad \qquad
    - \delta^{ij} G_{i} \dot{\alpha}_{j} 
    + 2 \delta^{ij} \dot{\tilde{V}}_{i} \tilde{V}_{j} 
    \,,\\
    &T^{\mu}{}_{i\mu} = 
    - a \dot{\alpha}_{i} 
    + \delta^{jk} \delta^{lm} \epsilon_{ijl} \partial_{k}{\tilde{V}_{m}} 
    - a \delta^{jk} \delta^{lm} \epsilon_{ijl} \alpha_{k} \dot{\tilde{V}}_{m} 
    + \delta^{jk} G_{j} \partial_{k}{\alpha_{i}} 
    \\
    & \qquad \qquad
    - \delta^{jk} G_{j} \partial_{i}{\alpha_{k}}
    - \delta^{jk} \alpha_{k} \partial_{i}{G_{j}} 
    + \delta^{jk} \tilde{V}_{i} \partial_{j}{\tilde{V}_{k}}
    + \delta^{jk} \tilde{V}_{j} \partial_{i}{\tilde{V}_{k}} 
    \,.
\end{split}
\label{VandpseudoVPs}
\end{equation}

\bibliography{Bibliography}
\bibliographystyle{utphys}

\end{document}